\documentclass{sig-alternate-05-2015}

\usepackage{cite}
\usepackage{graphicx}
\usepackage{algorithm}
\usepackage{algorithmic}
\usepackage{amsmath,amssymb,amsfonts}

\usepackage[textfont=normal,font=scriptsize,labelfont=bf]{subcaption}
\usepackage[font=footnotesize,labelfont=bf,textfont=bf]{caption}

\usepackage{amsmath}
\usepackage{comment}

\usepackage{verbatim,url}
\usepackage{color}

\definecolor{darkgreen}{RGB}{0,100,0}
\definecolor{darkred}{RGB}{139,0,0}
\definecolor{brown}{RGB}{210,105,30}

\usepackage{graphicx}
\usepackage{amsmath,amsfonts,amssymb}
\usepackage{multirow}
\usepackage{hhline}
\usepackage{verbatim, url}

\newcounter{theorem}
\newcounter{def}
\newcounter{mylea}
\newcounter{corollary}
\makeatletter
\renewcommand\@makefntext[1]{%
   \noindent\makebox[0em][r]{\@makefnmark}#1}
\makeatother

\vspace{-2.0in}

\allowdisplaybreaks

\begin{document}

\CopyrightYear{2016} 
\setcopyright{acmcopyright}
\conferenceinfo{ISPD '16,}{April 3--6, 2016, Santa Rosa, California, USA.}
\isbn{978-1-4503-4039-7/16/04}\acmPrice{\$15.00}
\doi{http://dx.doi.org/10.1145/2872334.2872361} 

\vspace{-2.0in}

\title{ePlace-3D: Electrostatics based Placement for 3D-ICs} 

\vspace{-4.0in}

\numberofauthors{5}
\author{
\alignauthor 
Jingwei Lu\\
\affaddr{Cadence Design Systems, Inc. } \\
\email{francesco.ljw@gmail.com}
\alignauthor 
Hao Zhuang\\
\affaddr{Dept. CSE of UCSD} \\
\email{hao.zhuang@cs.ucsd.edu}
\alignauthor 
Ilgweon Kang\\
\affaddr{Dept. CSE of UCSD} \\
\email{igkang@ucsd.edu}\\
\and
\alignauthor 
Pengwen Chen\\
\affaddr{Dept. Applied Maths of NCHU} \\
\email{pengwen@nchu.edu.tw}
\alignauthor 
Chung-Kuan Cheng\\
\affaddr{Dept. CSE of UCSD} \\
\email{ckcheng@ucsd.edu}
}	


\vspace{-2.0in}




\maketitle

\vspace{-0.2in}

\begin{abstract}
\label{sec:abs}
We propose a flat, analytic, mixed-size placement algorithm \textit{ePlace-3D} 
for three-dimension integrated circuits (3D-ICs) using nonlinear optimization.
Our contributions are 
(1) electrostatics based 3D density function with globally uniform smoothness 
(2) 3D numerical solution with improved spectral formulation 
(3) 3D nonlinear pre-conditioner for convergence acceleration
(4) interleaved 2D-3D placement for efficiency enhancement. 
Our placer outperforms the leading work
mPL6-3D and NTUplace3-3D with $6.44\%$ and $37.15\%$ shorter 
wirelength, $9.11\%$ and $10.27\%$ fewer 3D vertical interconnects (VI) 
on average of IBM-PLACE circuits.
Validation on the large-scale modern mixed-size (MMS) 
3D circuits shows high performance and scalability.
\end{abstract}


\vspace{-0.1in}

\section{Introduction}
\label{sec:intro}
Placement remains dominant on the overall quality  
of physical design automation~\cite{procha,miao_iccad13}. 
Based on logic synthesis~\cite{miao_iccad14},
back-end design on 
timing~\cite{shifter_icccas13},
power~\cite{zhang_slip14,wv_sim}, 
routability~\cite{cpre_todaes09,gr_slip11}, 
variability~\cite{tao_iccad13,chen_dac14} etc. 
are highly impacted by placement performance.
The emerging 3D-IC~\cite{luo_isvlsi12} 
challenges the traditional 2D
placers~\cite{eisen,fp3,complx,aplace3,mpl6,ntupl3,eplace} 
to produce 3D circuit layout 
with minimum wirelength yet 
limited vertical interconnects 
(through-silicon vias (TSVs), 
monolithic inter-tier vias (MIVs), etc.).
Innovations of mixed-size 3D-IC placement 
become quite desirable. 

Previous {\bf combinatorial} 3D-IC placers form two categories.
Folding based methods~\cite{cong07} folds
the 2D-IC placement layout to produce 3D solution 
with local refinement. 
Partitioning based approaches~\cite{goplen07,kim09} 
minimize the usage of vertical resources.
Kim et al.~\cite{kim09} partitions the netlist 
followed by tier assignment, 
then applies 2D quadratic placement~\cite{kw2} 
simultaneously over all the tiers.
{\bf Analytic} placers achieve better 3D-IC placement performace 
versus combinatorial algorithms. 
Goplen et al.~\cite{goplen03} models the 3D-IC placement 
by a quadratic framework~\cite{eisen}.
Hsu et al.~\cite{wa_tcad} extends the 2D-IC placement prototype~\cite{ntupl3-unified} 
and uses Bell-shape function~\cite{naylor} to smooth the vertical dimension. 
Luo et al.~\cite{mpl3d_tcad13} utilizes the 2D algorithm in~\cite{mpl6} 
and relaxes the discrete tiers via Huber function~\cite{huber}.
However, these modeling functions are only locally smooth.
Moreover, their hierarchical cell clustering and grid coarsening would degrade the quality~\cite{eplace}.
Separately, 
prior 3D placement benchmarks~\cite{ibm_place,iwls05} 
are of up to only 210K cells, which are too small to represent modern design complexity.
Large-scale bookshelf 3D-IC placement benchmarks 
become desirable.

\begin{figure*}
  \centering
  \begin{subfigure}[b]{0.24\textwidth}
    \centering
    \includegraphics[keepaspectratio, width=0.85\textwidth]{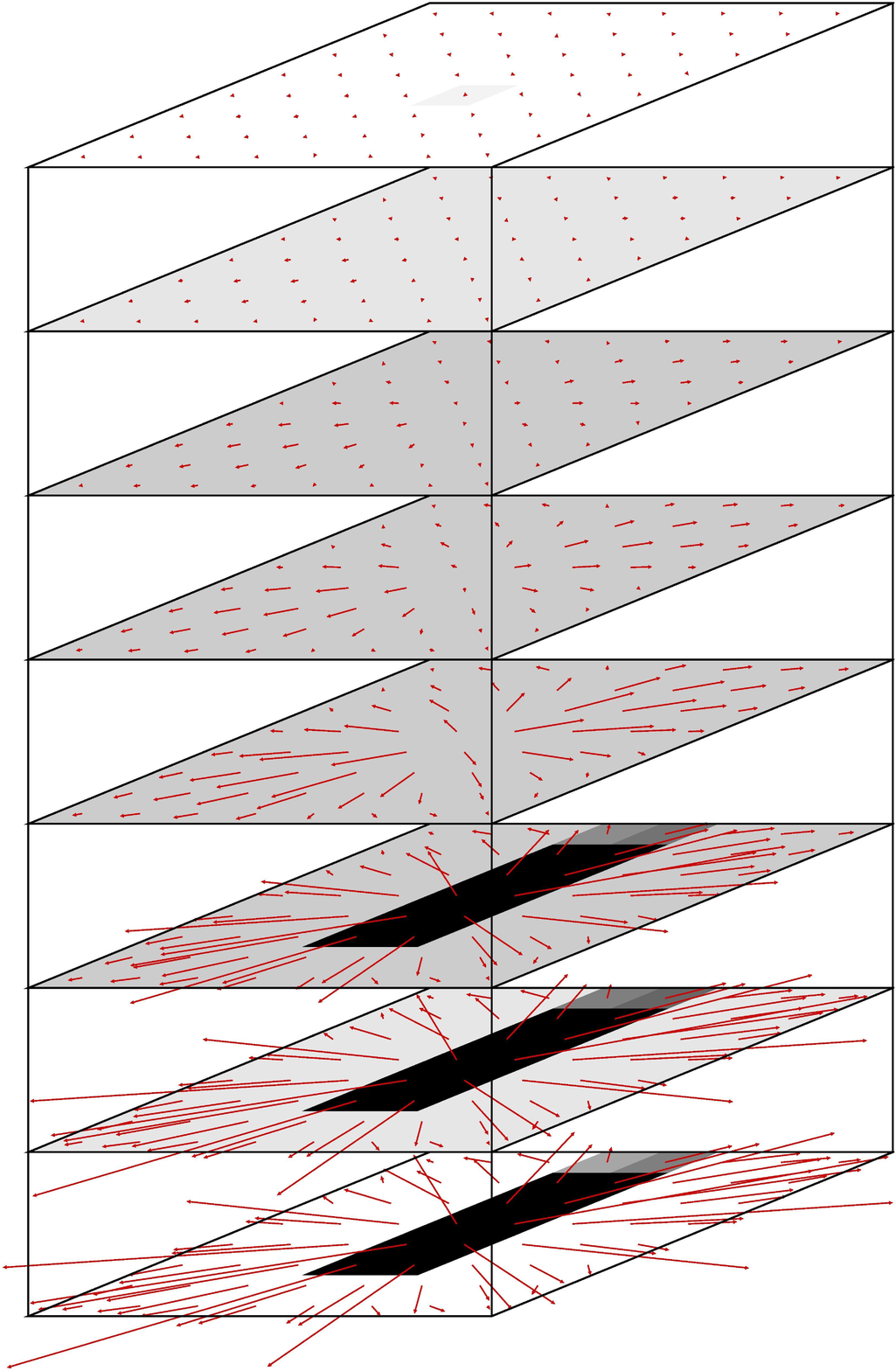}
    \caption{Iter=0, U=6.18e17, $\tau=90.13\%$.}
    \label{subfig:den}
  \end{subfigure}
  \begin{subfigure}[b]{0.24\textwidth}
    \centering
    \includegraphics[keepaspectratio, width=0.85\textwidth]{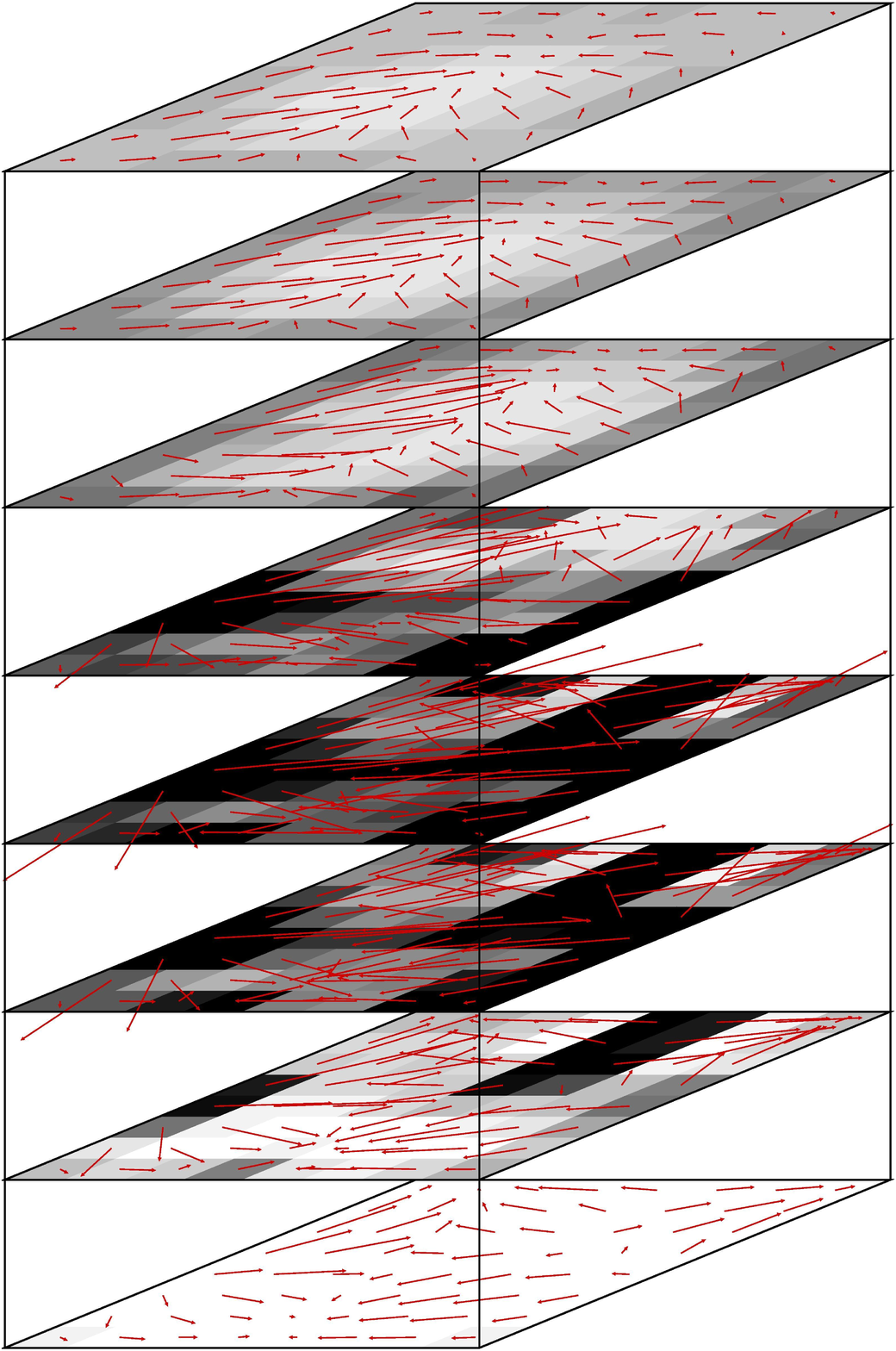}
    \caption{Iter=3, U=5.84e16, $\tau=47.56\%$.}
    \label{subfig:den}
  \end{subfigure}
  \begin{subfigure}[b]{0.24\textwidth}
    \centering
    \includegraphics[keepaspectratio, width=0.95\textwidth]{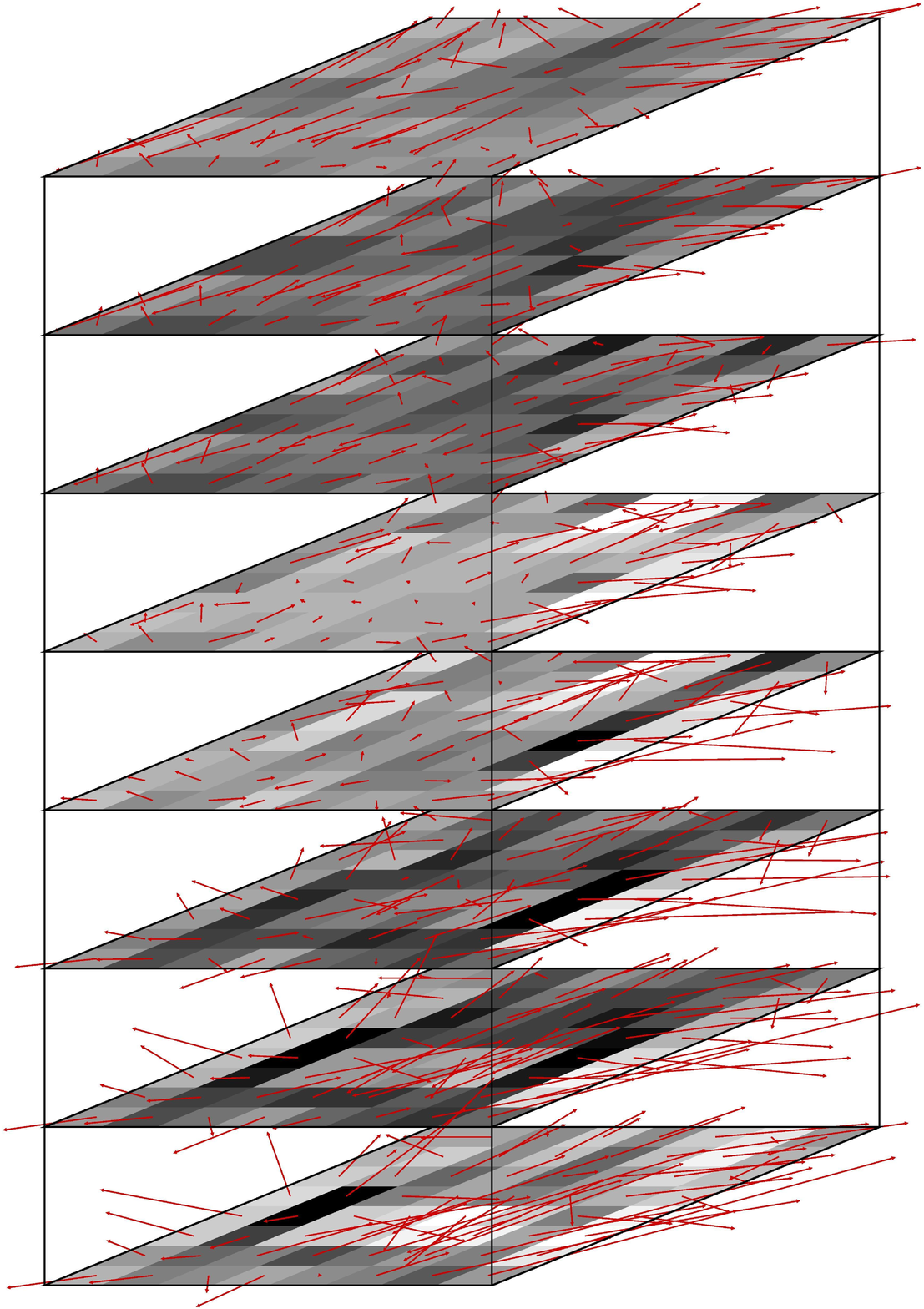}
    \caption{Iter=6, U=3.56e15, $\tau=11.75\%$.}
    \label{subfig:den}
  \end{subfigure}
  \begin{subfigure}[b]{0.24\textwidth}
    \centering
    \includegraphics[keepaspectratio, width=0.95\textwidth]{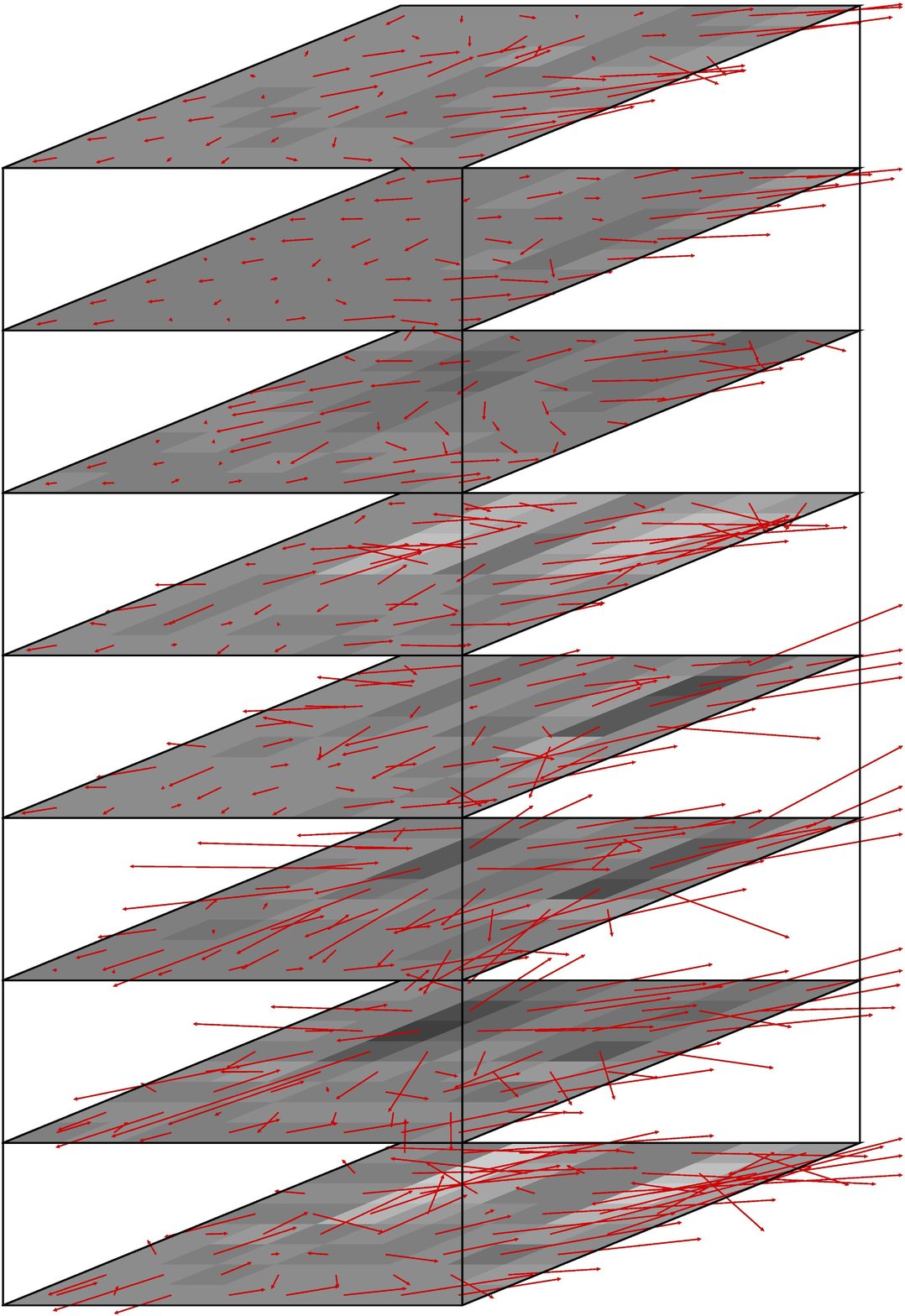}
    \caption{Iter=20, U=2.07e14, $\tau=2.53\%$.}
    \label{subfig:den}
  \end{subfigure}
  \caption{\bf Iterative density-driven global placement (wirelength force disabled)
  with potential $U$ and density overflow $\tau$ on the MMS ADAPTEC1 benchmark with three tiers 
  and resolution of $8\times 8\times 8$. 
  Electric density and field are shown by gray scale and red arrows. 
  All the movable objects are initialized at the bottom tier where 
  all IO blocks locate.
  {\bf eDensity-3D} iteratively spreads all the movable objects evenly within the entire 3D domain 
  to equalize the placement density.}  
  \label{fig:den}
\end{figure*}

{\bf In this work}, we extend the 2D placers ePlace~\cite{eplace,eplace-todaes,ljw_phd} 
and ePlace-MS~\cite{eplace-ms,ljw_phd} to the 3D domain. 
Our algorithm is named {\bf ePlace-3D} and focused on 
{\bf wirelength minimization} and {\bf density equalization}, 
while other 3D-IC objectives like thermal are not covered.
To the best of our knowledge, this is the first work in literature 
achieving analytically {\bf global smoothness} along all the three dimensions.
In contrast, previous analytic works~\cite{wa_tcad,mpl3d_tcad13} 
only ensure (partially) local smoothness in their density functions~\cite{naylor,huber}, 
while their less continuous cell movement 
would slow down placement convergence and cause
more penalty on wirelength.
We conduct analytic global placement and stochastic legalization
in the entire 3D cuboid domain, which maximizes the search space 
thus further boost the solution quality.
ePlace-3D well demonstrates the applicability of
the electrostatic density model {\bf eDensity}~\cite{fftpl,eplace-todaes} 
in various physical dimensions.
Our specific contributions are listed as follows.
\begin{itemize}
\item {\bf eDensity-3D}: an electrostatics based 3D density function 
ensuring global smoothness. 
\item A 3D numerical solution 
based on fast Fourier transform (FFT) and improved spectral formulation. 
\item A nonlinear 3D preconditioner to equalize 
all the moving objects in the optimization perspective. 
\item Interleaving coarse-grained 3D placement with 
fine-grained 2D placement to enhance efficiency.
\item Our mixed-size 3D-IC placement prototype {\bf ePlace-3D}
outperforms the leading placers mPL6-3D~\cite{mpl3d_tcad13} and 
NTUplace3-3D~\cite{wa_tcad} with 
$6.44\%$ and $37.15\%$ shorter wirelength, $9.11\%$ and $10.27\%$ 
fewer 3D vertical interconnects, while runs $2.55\times$ and $0.30\times$ faster
on average of all the ten IBM-PLACE benchmarks~\cite{ibm_place},
\end{itemize}
The remainder is organized as follows.
Section~\ref{sec:ana} introduces the background knowledge.
Section~\ref{sec:den} discusses our 3D placement 
density function eDensity-3D, numerical solution, and nonlinear 
precondition.
Section~\ref{sec:gp} provides an overview of ePlace-3D
algorithm.
Experiments and results are shown in Section~\ref{sec:exp}.
We conclude in Section~\ref{sec:conc}.

\section{Background}
\label{sec:ana}
Given a set $V$ of $n$ objects, 
net set $N$ and 3D cuboid core region 
$R=[0,d_x]\times[0,d_y]\times[0,d_z]$,  
global placement is formulated as constrained optimization.
The {\bf constraint} desires all the objects to be 
accommodated with zero overlap.
Let $\mathbf{v}$ denote the placement solution, which 
consists of the physical coordinates of all the objects.
The region $R$ is uniformly decomposed 
into $m_x\times m_y\times m_z$ 3D bins denoted as set $B$.
For every bin $b\in B$, 
the density $\rho_b(\mathbf{v})$ 
should not exceed the target density  
$\rho_t$. 
The {\bf objective} is to minimize the total 
half-perimeter wirelength (HPWL) of all the nets.
Let $HPWL_{e_x} = \max_{i,j\in e}|x_i-x_j|$ denote 
the horizontal wirelength of net $e$ (similar for 
$HPWL_{e_y}$), 
the total 2D HPWL is
$HPWL(\mathbf{v})=\sum_{e\in N}\left(\beta_xHPWL_{e_x}(\mathbf{v})+\beta_yHPWL_{e_y}(\mathbf{v})\right)$.
We use $\beta_x$, $\beta_y$ and $\beta_z$ as dimensional weighting factors.
3D-IC placement needs {\bf vertical interconnects}, such as 
through-silicon via (TSV) and 
monolithic inter-tier via (MIV), 
to penetrate silicon tiers. 
Diverse types of connects have different physical and electrical properties. 
However, ePlace-3D is compatible with any types of connects, which 
can be reflected on the weight of $\beta_z$. 
\textit{In the remainder of this manuscript, we name all types of 
3D vertical interconnects uniformly as $\mathbf{VI}$ for simplicity. 
The number of vertical interconnect units ($\#$VI) is computed as 
how many times silicon tiers have been penetrated, 
e.g., one vertical connect between tier one and tier three 
is counted as two $VI$.}
The nonlinear placement optimization is formulated as 
\begin{equation}
\begin{small}
\label{eq:gp}
\min_{\mathbf{v}}\text{ } \left(HPWL(\mathbf{v}) + \beta_z \#VI\right) \text{ s.t. } \rho_b(\mathbf{v})\le\rho_t\text{, }\forall b\in B.
\end{small}
\end{equation}

Analytic methods conduct placement using 
gradient-directed optimization.
As $HPWL(\mathbf{v})$ is not differentiable, 
we use {\bf wirelength smoothing} by 
weighted-average (WA) model~\cite{wa_tcad}.
\begin{equation}
\begin{small}
\label{eq:wa}
\begin{aligned}
W_{e_x}(\mathbf{v}) = & \frac{\sum_{i\in e}x_i\exp\left(x_i/\gamma_x\right)}{\sum_{i\in e}\exp\left(x_i/\gamma_x\right)} - \frac{\sum_{i\in e}x_i\exp\left(-x_i/\gamma_x\right)}{\sum_{i\in e}\exp\left(-x_i/\gamma_x\right)}
\end{aligned}
\end{small}
\end{equation}
Here $W_e(\mathbf{v})=\beta_xW_{e_x}(\mathbf{v})+
\beta_yW_{e_y}(\mathbf{v})+\beta_zW_{e_z}(\mathbf{v})$ 
and $W(\mathbf{v})=\sum_eW_e(\mathbf{v})$.
$\gamma_x$, $\gamma_y$ and $\gamma_z$ control the modeling accuracy.
{\bf Density function} relaxes all the $|B|$ constraints in 
Eq.~(\ref{eq:gp}).
Most 2D and 3D {\bf quadratic} placers~\cite{complx,kim09,goplen03} 
follow the linear density force formulation by~\cite{eisen}.
{\bf Nonlinear} placers~\cite{ntupl3-unified,mpl6,mpl3d_tcad13,wa_tcad} 
have their dedicated density functions.
NTUplace3-3D~\cite{wa_tcad} leverages bell-shape curve~\cite{naylor} for 
local smoothness in 3D domain.
mPL6-3D~\cite{mpl6} uses Helmholtz function to globally smoothen the 2D plane 
and Huber's function to locally smoothen the vertical dimension.
The {\bf electrostatics} based density function~\cite{eplace} converts 
objects to charges. By the Lorentz law, 
the electric repulsive force spreads charges away towards the 
electrostatic equilibrium 
state, which produces a globally even density distribution.
Let $U(\mathbf{v})$ denote the density cost function, 
the constraints in Eq.~(\ref{eq:gp}) can be relaxed by 
the penalty factor $\lambda$, while the unconstrained 
optimization is shown as below.
\begin{equation}
\label{eq:obj}
\min_{\mathbf{v}}f(\mathbf{v})=W(\mathbf{v})+\lambda U(\mathbf{v}),\\
\end{equation}

In this work, we set vertical connects as zero-volumed thus do not consider them 
in eDensity-3D\footnote{Practically, vertical connects can never
be zero volumed. However, for academic research we are able to 
simplify the engineering problems to boost scientific innovations.
Similarly, state-of-the-art 2D placement academic
works~\cite{eplace,mpl6,aplace3,ntupl3,complx,fp3,kw2} 
target wirelength only and ignore other objectives like 
timing, power and routability.
As vertical connects may be of large volume thus significantly 
contribute to the placement density, 
we will put it in our future work.}.
Therefore, the optimization of electrostatics will not be affected 
and can be still achieved based on the movement of netlist objects.
{\bf Density overflow} is used to terminate global placement and 
denoted as $\tau$, which is 
\begin{equation}
\tau = \frac{\sum_{b\in B} \max\left(V^{m}_b-\rho_t V^{WS}_b,0\right)}{V_m}.
\end{equation}
Here $V_{m}$ is the total volume of all the movable objects,  
$V^{m}_b$ is the total volume of objects in the 
bin $b$, and $V^{WS}_b$ is the total whitespace in bin $b$.
The volume of each cell is computed as its planar area multiplied by 
the depth of each tier. 


\section{eDensity-3D: 3D Density Function}
\label{sec:den}

In this section, we introduce our novel 3D
density function {\bf eDensity-3D}, 
a fast numerical solution by spectral methods,
and approximated 3D nonlinear preconditioner. 
The key insight is, 
\textit{we treat the third dimension equally as the other two 
dimensions, such that vertical cell movement 
will be as smooth as the planar movement in 2D placement.}
The behavior of eDensity-3D is visualized in Figure~\ref{fig:den}.


\subsection{3D Density Function}
\label{subsec:den}

Extending the planar function eDensity in~\cite{eplace},
eDensity-3D models the entire placement instance as a 3D electrostatic field.
Every placement object (standard cells, macros and fillers)
is converted to a positively charged cuboid.
The electric repulsive force spreads all
the objects away from the high-density region. 
The 3D density cost $U$ is 
modeled as the total potential energy of the system 
and defined as below
\begin{equation}
\begin{scriptsize}
\label{eq:potn}
U(\mathbf{v})=\sum_{i\in V}U_i(\mathbf{v})=\sum_{i\in V}q_i\Phi_i(\mathbf{v}).
\end{scriptsize}
\end{equation}
$q_i$ denotes the electric quantity of the charge $i$ 
and is set as the physical volume of 
placement object $i$.
$\Phi_i$ is the electric potential at charge $i$. 
Charges with high potential will  
reduce the placement overlap by moving towards 
the direction of largest energy descent.
Unlike the spatial density distribution $\rho(x,y,z)$ (Figure~\ref{fig:den})
which is coarse and non-differentiable, 
the electric potential distribution $\Phi(x,y,z)$ is globally smooth. 
We use the potential gradient (thus electric field), $\nabla\Phi(x,y,z)=\mathbf{E}(x,y,z)$, 
to direct cell movement for density equalization.
Given a placement layout $\mathbf{v}$, 
we generate the density map $\rho(x,y,z)$, 
then compute the potential map $\Phi(x,y,z)$ 
by solving the {\bf 3D Poisson's equation}
\begin{equation}
\begin{small}
\label{eq:poi}
\begin{cases}
\nabla\cdot\nabla\Phi (x,y,z) = -\rho(x,y,z)\text{,} \\
\mathbf{\hat{n}}\cdot\nabla\Phi (x,y,z) = 0 \text{,     } (x,y,z)\in \partial R, \\
\iiint\limits_R \Phi(x,y,z) = \iiint\limits_R\rho(x,y,z)=0\text{.} \\
\end{cases}
\end{small}
\end{equation}
Here $\hat{n}$ is the outer unit normal of the 
placement cube $R$. 
$\partial R$ is the boundary and consists of 
{\textit orthogonal rectangular planes} 
to enclose the placement cuboid.
In Eq.~(\ref{eq:poi}), 
the first equation has 
$\nabla\cdot\nabla\equiv\frac{\partial^2}{\partial x^2}+\frac{\partial^2}{\partial y^2}+\frac{\partial^2}{\partial z^2}$.
Neumann condition by the second equation
requires that when any object $i$ reaches any boundary plane, 
its density force vector will have the
component perpendicular to the plane reduced to zero, 
in order to prevent $i$ from penetrating the plane.
The third equation shows that 
the integral of density $\rho(x,y,z)$ and potential $\Phi(x,y,z)$ 
within $R$ are set to zero to ensure that 
(1) electric force drives all the charges towards even 
density distribution rather than pushing them to infinity, 
which matches the placement objective
(2) the 3D Poisson's equation would have a unique solution 
by satisfying the Neumann condition.
We differentiate the potential $\Phi_i$ on each charge $i$
to generate the electric field 
$\nabla\Phi_i=\mathbf{E_i}=\left(E_{i_x},E_{i_y},E_{i_z}\right)$.
The electric (density) force is $\nabla U_i=q_i\nabla\Phi_i=q_i\mathbf{E_i}$.

\subsection{3D Numerical Solution}
\label{subsec:num}

Based on the 2D solution in~\cite{eplace}, 
we solve the 3D Poisson's equation
by spectral methods using frequency decomposition~\cite{fft_poi}.
To satisfy the Neumann condition of zero gradients at the boundaries, 
we use sinusoidal wave to express the electric field $\mathbf{E}(x,y,z)$.
We construct an odd and periodic field distribution 
by negatively mirroring itself w.r.t. the origin, 
then periodically extending it towards positive and negative infinities.
Electric potential and density distributions are then expressed 
by cosine waveforms, which are the integration and 
differentiation of the field. 
Let $a_{j,k,l}$ denote the 3D coefficients of the density frequency. 
\begin{equation}
\begin{small}
\label{eq:fft_coef}
\begin{aligned}
a_{j,k,l}=\frac{1}{n^3}\sum_{x,y,z}\rho(x,y,z)\cos(w_jx)\cos(w_ky)\cos(w_lz)
\end{aligned}
\end{small}
\end{equation}

eDensity~\cite{eplace} sets $w_j=\pi\frac{j}{m_x}$, which equals 
the discrete index for the $j$th frequency component. 
However, as we are conducting placement in a continuous 
domain, the multiplication of $x$ and $w_j$ induces
inconsistency. 
In this work, we propose {\bf improved spectral methods} for 
the 3D placement density function. 
Specifically, we set 
$w_j=\pi\frac{j}{d_x}$ since $x$ ranges within $(0, d_x)$. 
As a result, $w_jx=\pi j\frac{x}{d_x}$
well matches the original unit of discrete frequency index, 
and we have all the frequency indexes defined as 
$\lbrace w_j,w_k,w_l\rbrace=\lbrace\frac{\pi j}{d_x}, \frac{\pi k}{d_y}, \frac{\pi l}{d_z}\rbrace$.
As mentioned in Section~\ref{sec:ana}, 
here $d_x$, $d_y$ and $d_z$ represent the dimensions of the 
cuboid placement core region.
$d_{\lbrace x,y,z\rbrace}$ can be set as any value 
since $w_{\lbrace j,k,l\rbrace}$ will be normalized by 
$\frac{\lbrace x,y,z\rbrace}{d_{\lbrace x,y,z\rbrace}}$.
$j$, $k$ and $l$ range in $[0,n-1]$, 
which is only half of a cosine function period. 
In contrast, one complete function period 
centered at the origin is $[-n,n-1]$.
Therefore, we have $\pi$ rather than 
$2\pi$ in the above frequency index.
We set $a_{0,0,0}=0$ to remove the zero-frequency component.
The spatial density distribution $\rho(x,y,z)$ is 
\begin{equation}
\begin{small}
\label{eq:fft_den}
\begin{aligned}
\rho(x,y,z) = & \sum_{j,k,l}a_{j,k,l}\cos(w_jx)\cos(w_ky)\cos(w_lz)\text{.} \\
\end{aligned}
\end{small}
\end{equation}

\begin{figure}[http]
\centering
\includegraphics[width=1.0\columnwidth, angle=0]{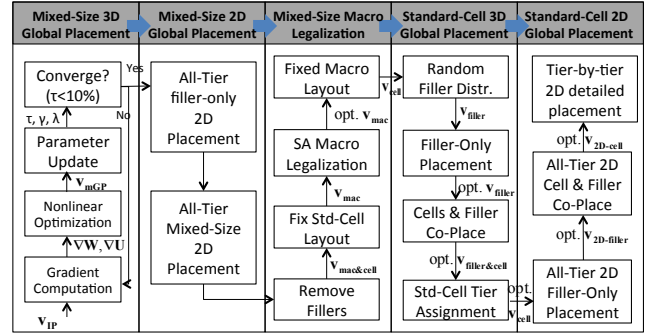}
\caption{the flowchart of ePlace-3d.}
\label{fig:pl}
\end{figure}

To achieve $\nabla\cdot\nabla\Phi(x,y,z)=-\rho(x,y,z)$, 
the solution to the potential can be expressed as 
\begin{equation}
\begin{scriptsize}
\label{eq:fft_potn}
\begin{aligned}
\Phi(x,y,z) = \sum_{j,k,l}\frac{a_{j,k,l}}{w_j^2+w_k^2+w_l^2}\cos(w_jx)\cos(w_ky)\cos(w_lz).
\end{aligned}
\end{scriptsize}
\end{equation}
By differentiating Eq.~(\ref{eq:fft_potn}), 
we have the electric field distribution 
$\mathbf{E}(x,y,z)=(E_x,E_y,E_z)$ shown as below
\begin{equation}
\begin{small}
\label{eq:fft_field}
\begin{aligned}
\begin{cases}
E_x (x,y,z) = \sum_{j,k,l}\frac{a_{j,k,l}w_j}{w_j^2+w_k^2+w_l^2}\sin(w_jx)
\cos(w_ky)\cos(w_lz), \\
E_y (x,y,z) = \sum_{j,k,l}\frac{a_{j,k,l}w_k}{w_j^2+w_k^2+w_l^2}\cos(w_jx)
\sin(w_ky)\cos(w_lz), \\
E_z (x,y,z) = \sum_{j,k,l}\frac{a_{j,k,l}w_l}{w_j^2+w_k^2+w_l^2}\cos(w_jx)
\cos(w_ky)\sin(w_lz). \\
\end{cases}
\end{aligned}
\end{small}
\end{equation}
Let $|B|=m_x\times m_y\times m_z$ denote the 
total number of bins in global placement.
Instead of quadratic complexity, 
above spectral equations can be efficiently 
solved using FFT algorithms~\cite{fft} with 
$O\left(|B| log |B|\right)$ complexity.

\subsection{3D Nonlinear Precondition}
\label{subsec:pre}

Theoretically, preconditioning improves 
convergence rate rather than solution quality. 
However, as placement is a highly nonlinear, 
non-convex and ill-conditioned problem,
the Hessian matrix with improved condition number
would reshape the search direction for the nonlinear 
solver to follow.
As a result, preconditioning would open the gate 
for unexplored high-dimension search space, while
surprising quality enhancement would be expectable.

Preconditioned mixed-size placement should tolerate the huge
physical and topological differences between all the 
standard cells, macros, and dummy fillers.
In~\cite{eplace}, the nonlinear preconditioner $H$ for 
2D placement is modeled as
\begin{equation}
\begin{small}
\label{eq:precon_2d}
H_{i_x}=\frac{\partial^2 f}{\partial x_i^2}=\frac{\partial^2 W}{\partial x_i^2} + 
\lambda\frac{\partial^2 U}{\partial x_i^2}\approx |N_i|+\lambda A_i.
\end{small}
\end{equation}
Here $N_i$ are all the nets incident to the object $i$, 
$A_i$ is the 2D area of the object $i$. In 3D placement, 
we use $V_i$ to denote the volume of $i$ instead.
The preconditioned gradient $\nabla f_{pre}=H^{-1}\nabla f$ then
improves and accelerates the placement.
Our studies show that 
Eq.~(\ref{eq:precon_2d}) relies on the assumption of 
$\frac{\partial^2 W/\partial x_i^2}{\partial^2 U/\partial x_i^2}\approx\frac{|N_i|}{A_i}$.
However, the third dimension weakens $\frac{\partial^2 W}{\partial x_i^2}$ and 
breaks the above assumption. 
As a result, $|N_i|$ dominates $\lambda V_i$ 
and makes fillers and macros with small $|N_i|$ spread
faster than standard cells, as Eq.~(\ref{eq:precon_3d_ineq}) shows
\begin{equation}
\begin{scriptsize}
\label{eq:precon_3d_ineq}
\left\Vert H_{i_x}^{-1}\nabla U\right\Vert=
\left\Vert\frac{\partial U/\partial x_i}{\lambda V_i}\right\Vert \gg 
\left\Vert\frac{\partial f/\partial x_i}{|N_i|+\lambda V_i}\right\Vert = 
\left\Vert H_{i_x}^{-1}\nabla f\right\Vert.
\end{scriptsize}
\end{equation}
Instead, we propose a new preconditioner as below
\begin{equation}
\begin{small}
\label{eq:precon_3d}
H_{i_x}=\frac{\partial^2 f}{\partial x_i^2}\approx \lambda\frac{\partial^2 U}{\partial x_i^2}\approx \lambda V_i,
\end{small}
\end{equation}
The noise factors introduced by $|N_i|$ is resolved, 
where all the objects are being equalized in the optimizer's 
perspective and simultaneously spread over the entire 
domain.
Experiments show that our 3D preconditioner 
reduces the global placement iterations 
by $15\%$ and improves the wirelength by $30\%$ 
over all the 16 MMS benchmarks.

\subsection{Complexity}
\label{subsec:bal}
{\bf Complexity} significantly impacts the placement runtime.
In each iteration, we traverse all the bins to reset their 
density in $O(|B|)$ time, 
then traverse all the placement objects in $O(n)$ time to update 
the superimposed density map.
By Eq.~(\ref{eq:fft_coef}), (\ref{eq:fft_potn}) and (\ref{eq:fft_field}),
five times of 3D FFT computation are invoked, 
which costs $O(5n\log n)$ time. 
By our grid sizing 
strategy in Eq.~(\ref{eq:grid_sz}), 
$|B|/n$ is limited to constant.
The overall complexity is thus
$O(|B|+n+5n\log n)\approx O(n\log n)$, 

In ePlace-3D, 
the placement domain 
is geometrically transformed from 
$R=[0,d_x]\times[0,d_y]\times[0,d_z]$ to
$R'=[0,1]\times[0,1]\times[0,1]$. 
We set the density resolutions 
$m_x=m_y=m_z=m_{3D}$ to make the placement domain $R'$ 
uniformly decomposed into $|B|=m_{3D}^3$ cubic bins.
Let $V_R$ denote the total volume of $R$ and $V_{C_{avg}}$ denote the 
average area of all standard cells.
The grid sizing is set as 
\begin{equation}
\begin{scriptsize}
\label{eq:grid_sz}
|B|=\frac{V_R}{k\times V_{C_{avg}}\times \rho_t^{-1}}.
\end{scriptsize}
\end{equation}
Here every $k$ standard cells are accommodated by one bin. 
Placement quality (efficiency) is determined by the value of $k$.
In this work, we constantly set $k=1.0$.


\newpage

\section{ePlace-3D: Overview}
\label{sec:gp}

ePlace-3D is built upon
the infrastructure of ePlace-MS~\cite{eplace-ms}.
Figure~\ref{fig:pl} shows the flowchart of our algorithm. 
Given a placement instance, 
our algorithm minimizes the quadratic 
wirelength over the 3D domain
to produce the initial solution $\mathbf{v_{IP}}$.
To approach the optimum solution in the end, 
we make $\mathbf{v_{IP}}$ as
minimum-wirelength violation-tolerant.

\begin{figure}[http]
  \centering
  \begin{subfigure}[b]{0.22\textwidth}
    \centering
    \includegraphics[keepaspectratio, width=0.95\textwidth]{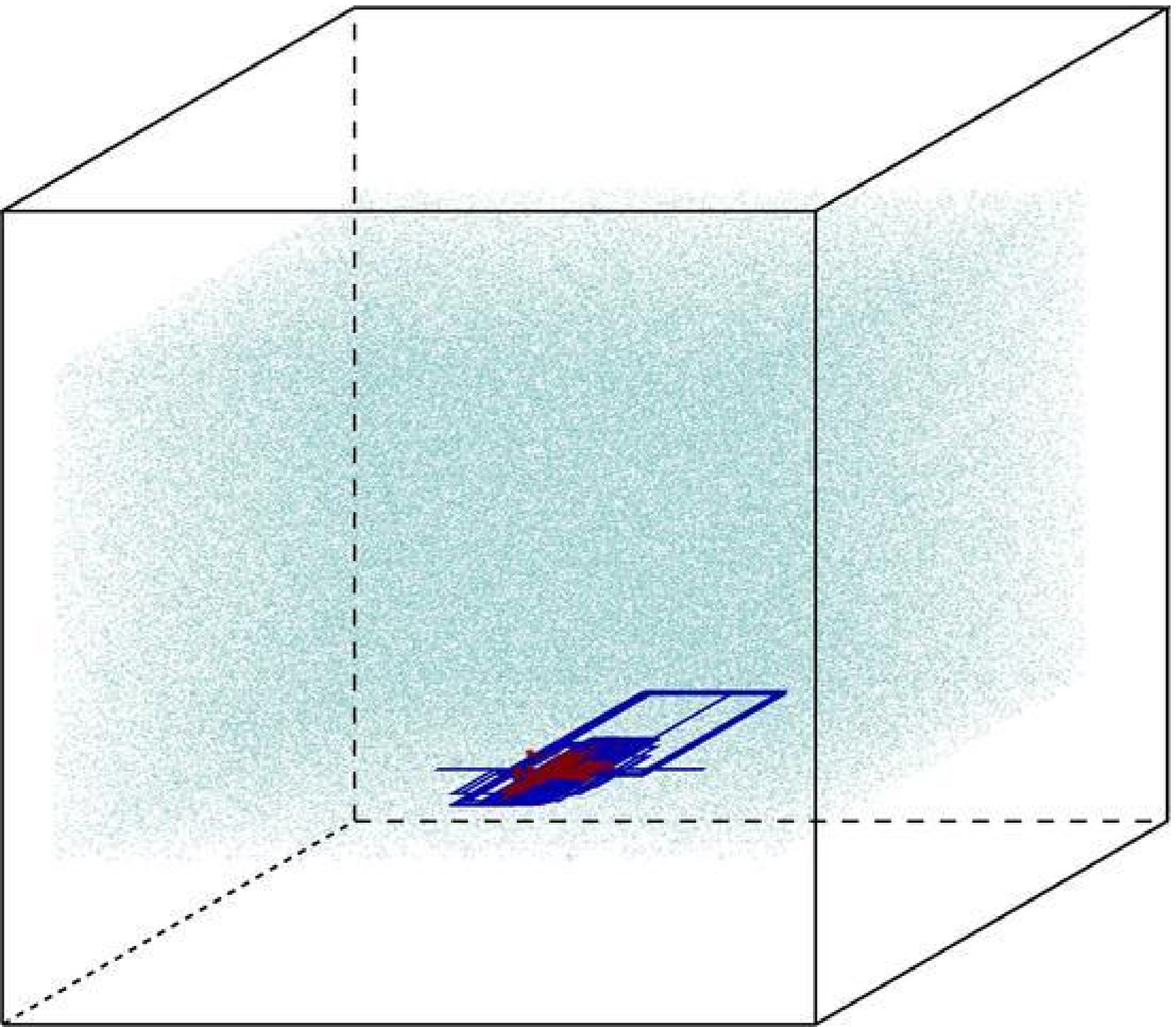}
    \caption{Iter=0, WL=1.32e7, $\#$VI=0, $\tau=93.7\%$.}
    \label{subfig:den}
  \end{subfigure}
  \begin{subfigure}[b]{0.22\textwidth}
    \centering
    \includegraphics[keepaspectratio, width=0.95\textwidth]{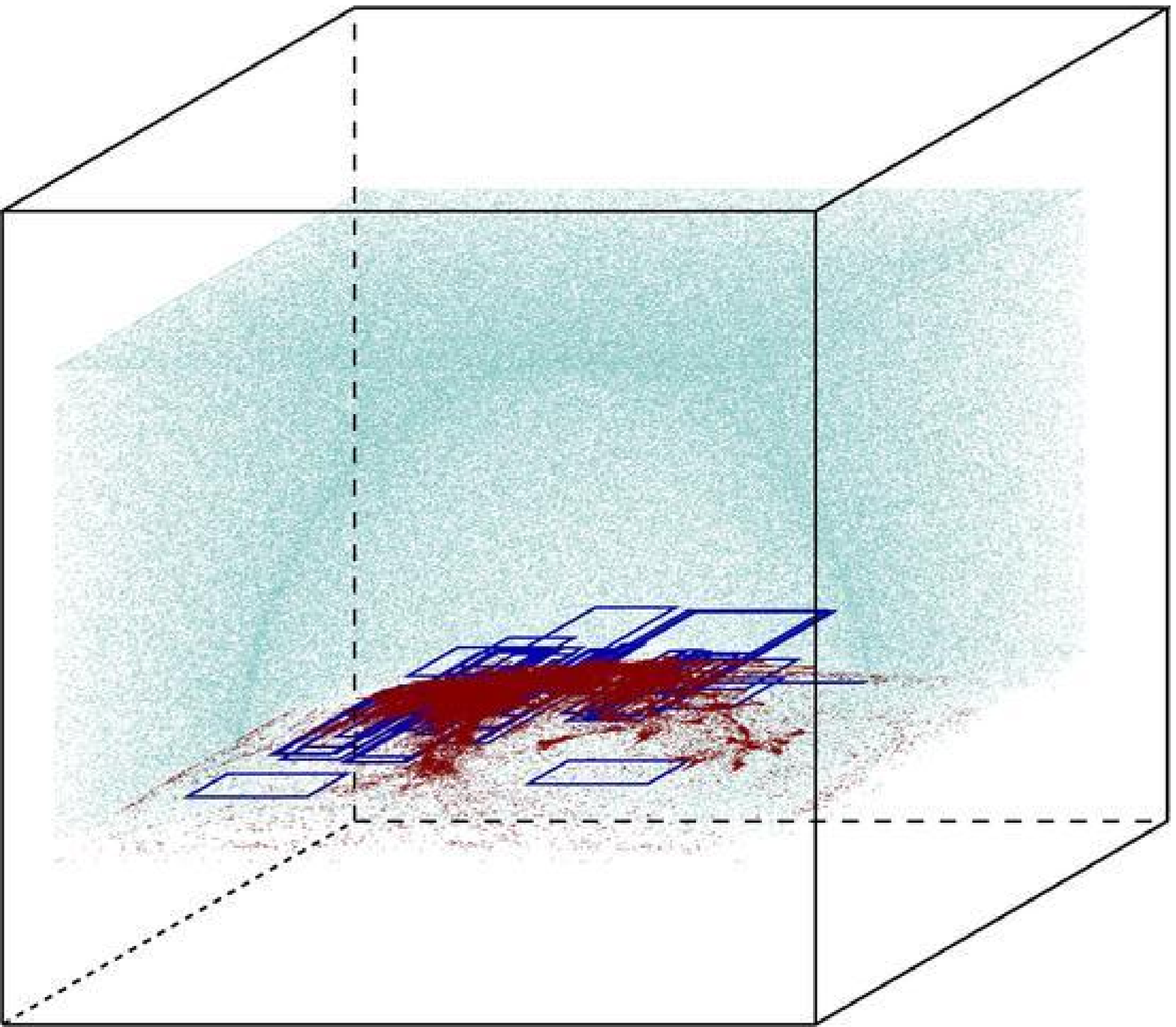}
    \caption{Iter=266, WL=3.29e7, $\#$VI=1.35e3, $\tau=77.1\%$.}
    \label{subfig:den}
  \end{subfigure}
  \begin{subfigure}[b]{0.22\textwidth}
    \centering
    \includegraphics[keepaspectratio, width=0.95\textwidth]{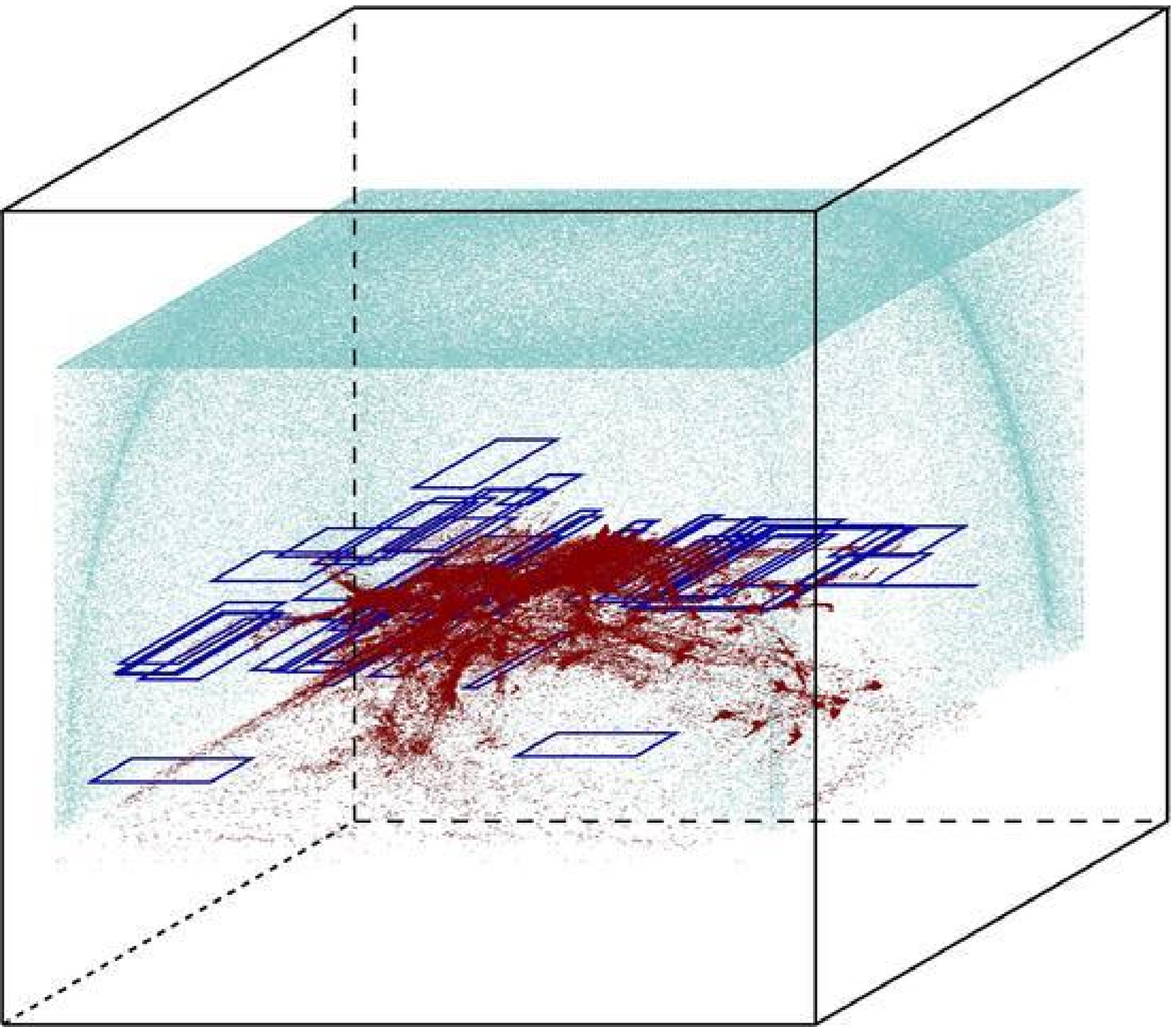}
    \caption{Iter=328, WL=3.91e7, $\#$VI=4.14e3, $\tau=61.1\%$.}
    \label{subfig:den}
  \end{subfigure}
  \begin{subfigure}[b]{0.22\textwidth}
    \centering
    \includegraphics[keepaspectratio, width=0.95\textwidth]{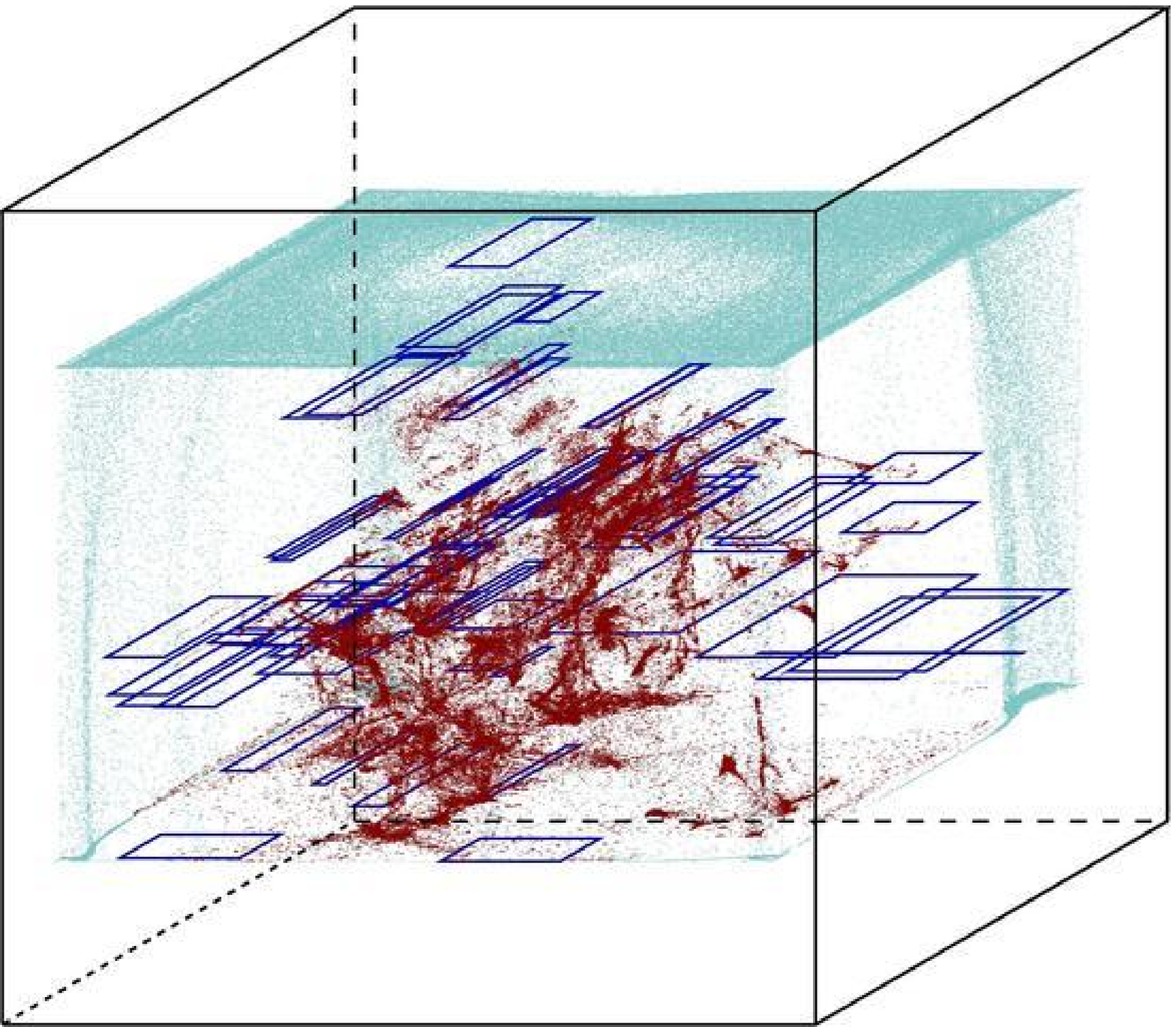}
    \caption{Iter=376, WL=4.21e7, $\#$VI=7.70e3, $\tau=45.2\%$.}
    \label{subfig:den}
  \end{subfigure}
  \begin{subfigure}[b]{0.22\textwidth}
    \centering
    \includegraphics[keepaspectratio, width=0.95\textwidth]{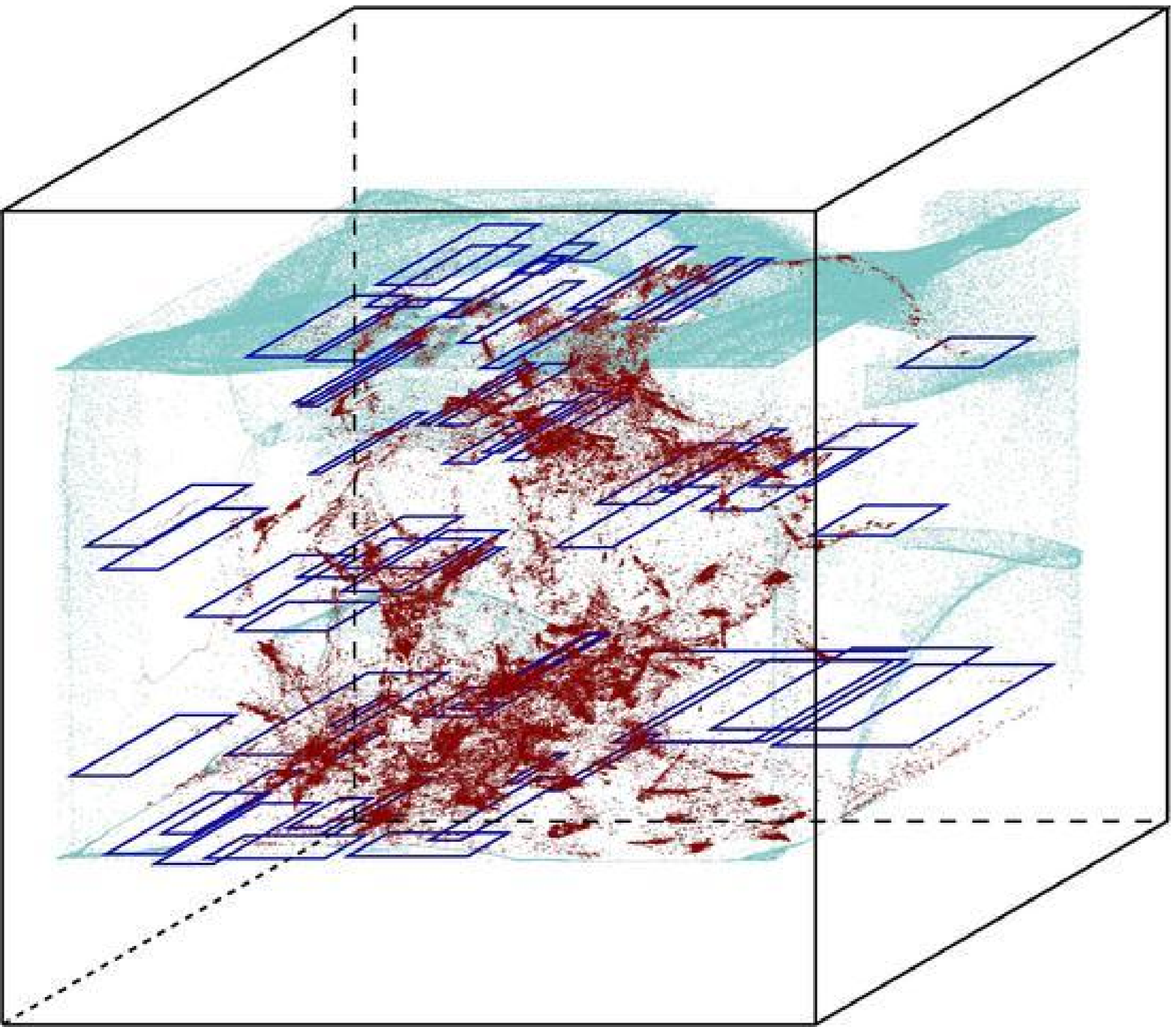}
    \caption{Iter=432, WL=4.64e7, $\#$VI=8.57e3, $\tau=28.9\%$.}
    \label{subfig:den}
  \end{subfigure}
  \begin{subfigure}[b]{0.22\textwidth}
    \centering
    \includegraphics[keepaspectratio, width=0.95\textwidth]{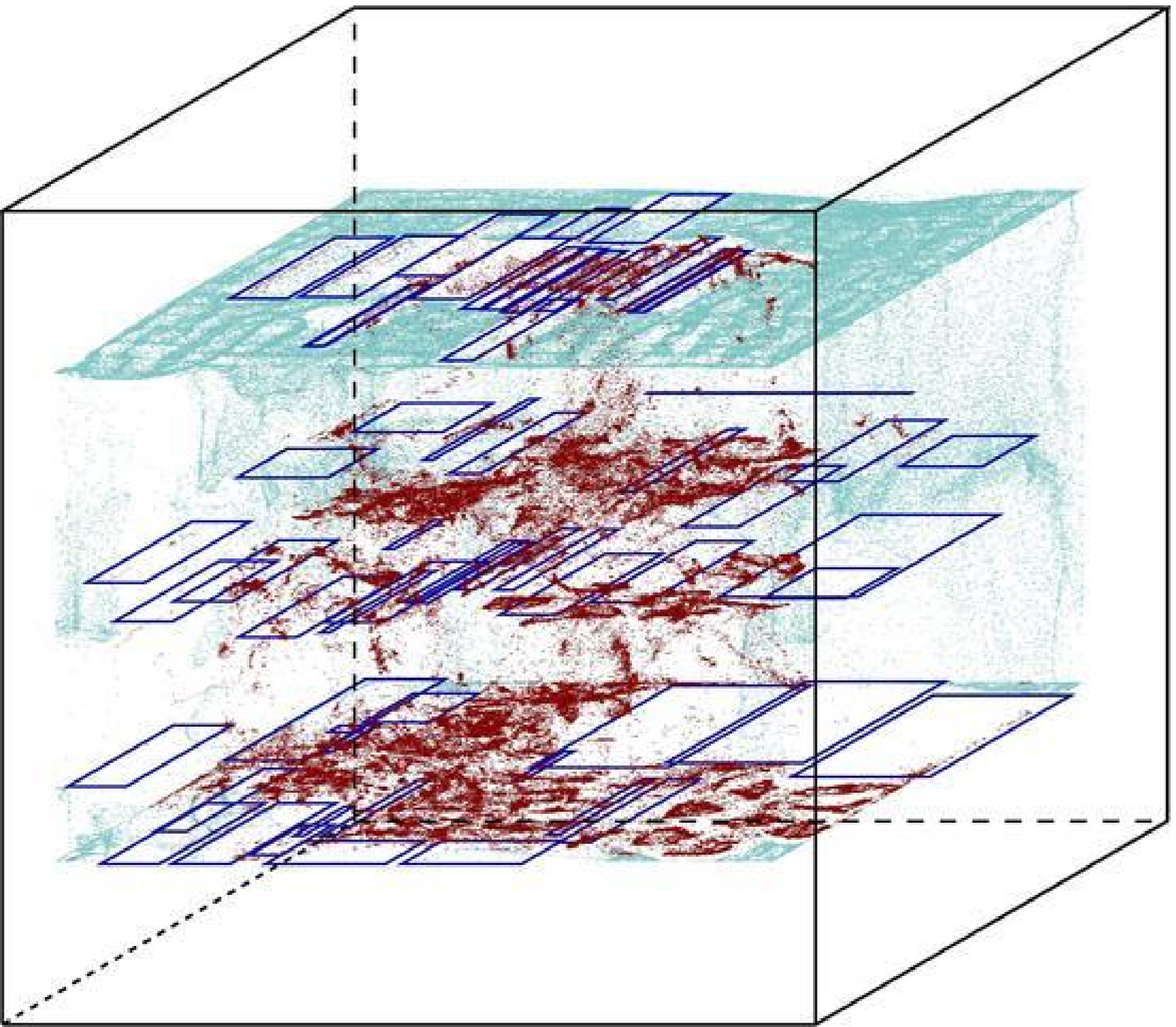}
    \caption{Iter=481, WL=5.06e7, $\#$VI=8.70e3, $\tau=14.9\%$.}
    \label{subfig:den}
  \end{subfigure}
  \caption{\bf 3D-IC mixed-size global placement on MMS 
  ADAPTEC1 with three tiers. Standard cells, macros and fillers are denoted by red dots, 
  blue rectangles and cyan dots.}  
  \label{fig:mgp3d}
\end{figure} 

Our 3D-IC global placement is visualized in Figure~\ref{fig:mgp3d}. 
Unconnected fillers~\cite{mpl6,eplace} are inserted 
to populate up extra whitespace.
All the fillers are equally sized by the average 
dimensions of all the standard cells.
\textit{The optimum solution of 3D global placement will have 
all the cells, macros and fillers orient towards discrete tiers.}
Otherwise, some cuboid placement sites will be partially wasted,
degrading the solution quality. 
Figure~\ref{fig:mgp3d}(f) illustrates the beauty of our approach, 
i.e., the analytic 3D placer is visually approaching density evenness 
from the vertical dimension, 
which ensures \textit{negligible quality overhead during tier assignment}.
We use Nesterov's method~\cite{nesterov} as the nonlinear solver 
and determine the steplength by~\cite{eplace}.

\begin{figure}[h]
  \centering
  \begin{subfigure}[b]{0.22\textwidth}
    \centering
    \includegraphics[keepaspectratio, width=0.95\textwidth]{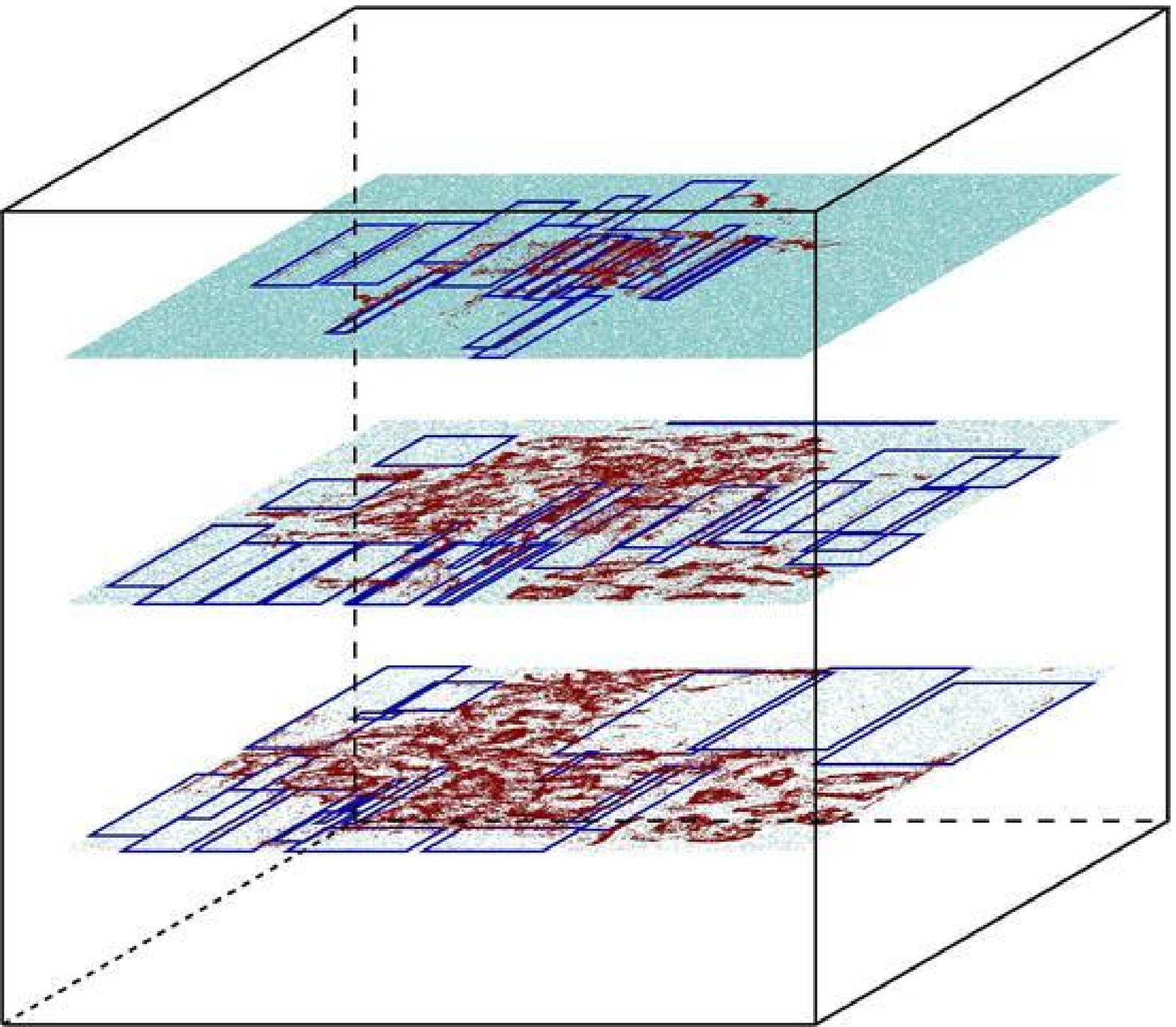}
    \caption{Iter=0, WL=4.71e7, $\#$VI=7.96e3, $\tau=31.9\%$.}
    \label{subfig:den}
  \end{subfigure}
  \begin{subfigure}[b]{0.22\textwidth}
    \centering
    \includegraphics[keepaspectratio, width=0.95\textwidth]{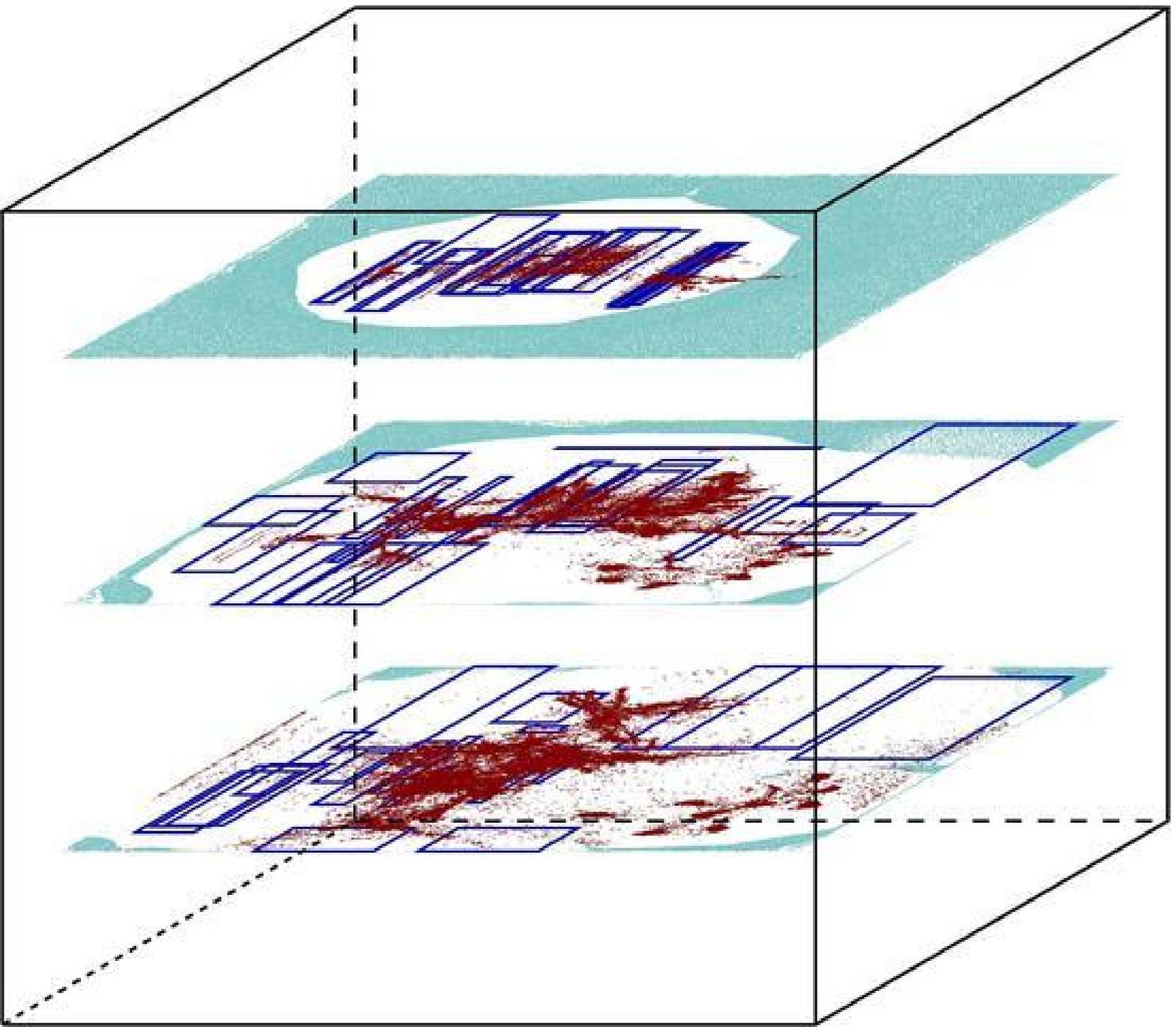}
    \caption{Iter=246, WL=3.64e7, $\#$VI=7.96e3, $\tau=50.3\%$.}
    \label{subfig:den}
  \end{subfigure}
  \begin{subfigure}[b]{0.22\textwidth}
    \centering
    \includegraphics[keepaspectratio, width=0.95\textwidth]{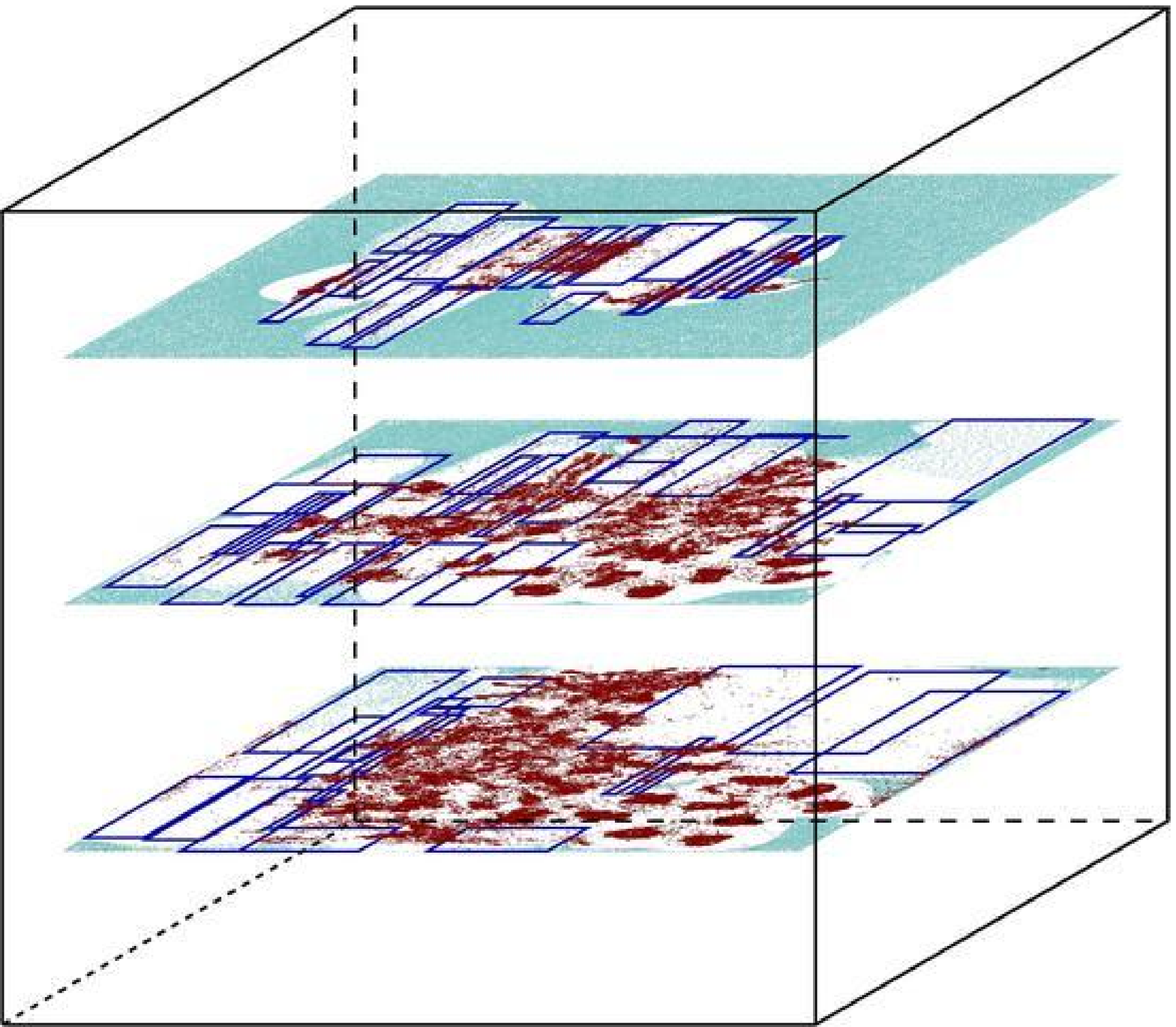}
    \caption{Iter=322, WL=4.46e7, $\#$VI=7.96e3, $\tau=31.9\%$.}
    \label{subfig:den}
  \end{subfigure}
  \begin{subfigure}[b]{0.22\textwidth}
    \centering
    \includegraphics[keepaspectratio, width=0.95\textwidth]{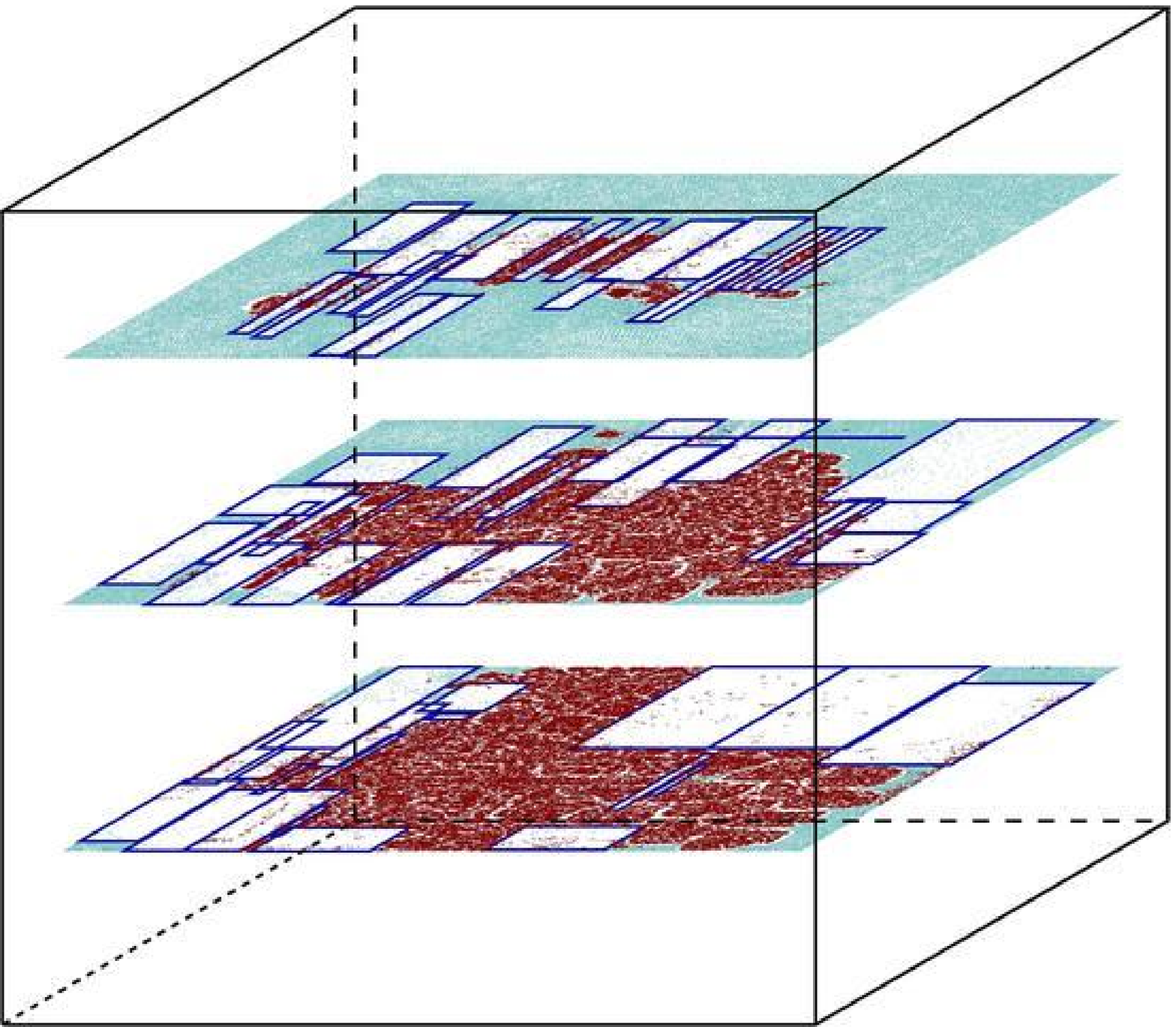}
    \caption{Iter=395, WL=4.99e7, $\#$VI=7.96e3, $\tau=14.8\%$.}
    \label{subfig:den}
  \end{subfigure}
  \caption{2D-IC mixed-size global placement on MMS 
  ADAPTEC1 with three tiers. 
  Initial and final overflow are both larger 
  than the final overflow of 3D placement 
  due to finer granularity ($m_{2D} \gg m_{3D}$).}  
  \label{fig:mgp2d}
\end{figure}

A multi-tier 2D-IC mixed-size global placement 
follows by assigning all the macros and standard cells 
to the closest tiers and separately filling the remaining 
whitespace on each tier with fillers.
Planar placement is conducted simultaneously over all the tiers. 
As wirelength smoothing is homogeneous over all the tiers 
(with the same $\gamma$), 
heterogeneous grid sizing is not feasible as density 
force is dependent on resolution by Eq.~(\ref{eq:fft_field}).
We set all the tiers with the same density resolution $m_{2D}$,
which is the maximum of that of all the tiers by Eq.~(\ref{eq:grid_sz}) with $k=1$.
In practice, we have $O(m^2_{2D})\approx O(m_{3D}^3)$.
Figure~\ref{fig:mgp2d} illustrates the progression. 

\begin{figure}[h]
  \centering
  \begin{subfigure}[b]{0.22\textwidth}
    \centering
    \includegraphics[keepaspectratio, width=0.95\textwidth]{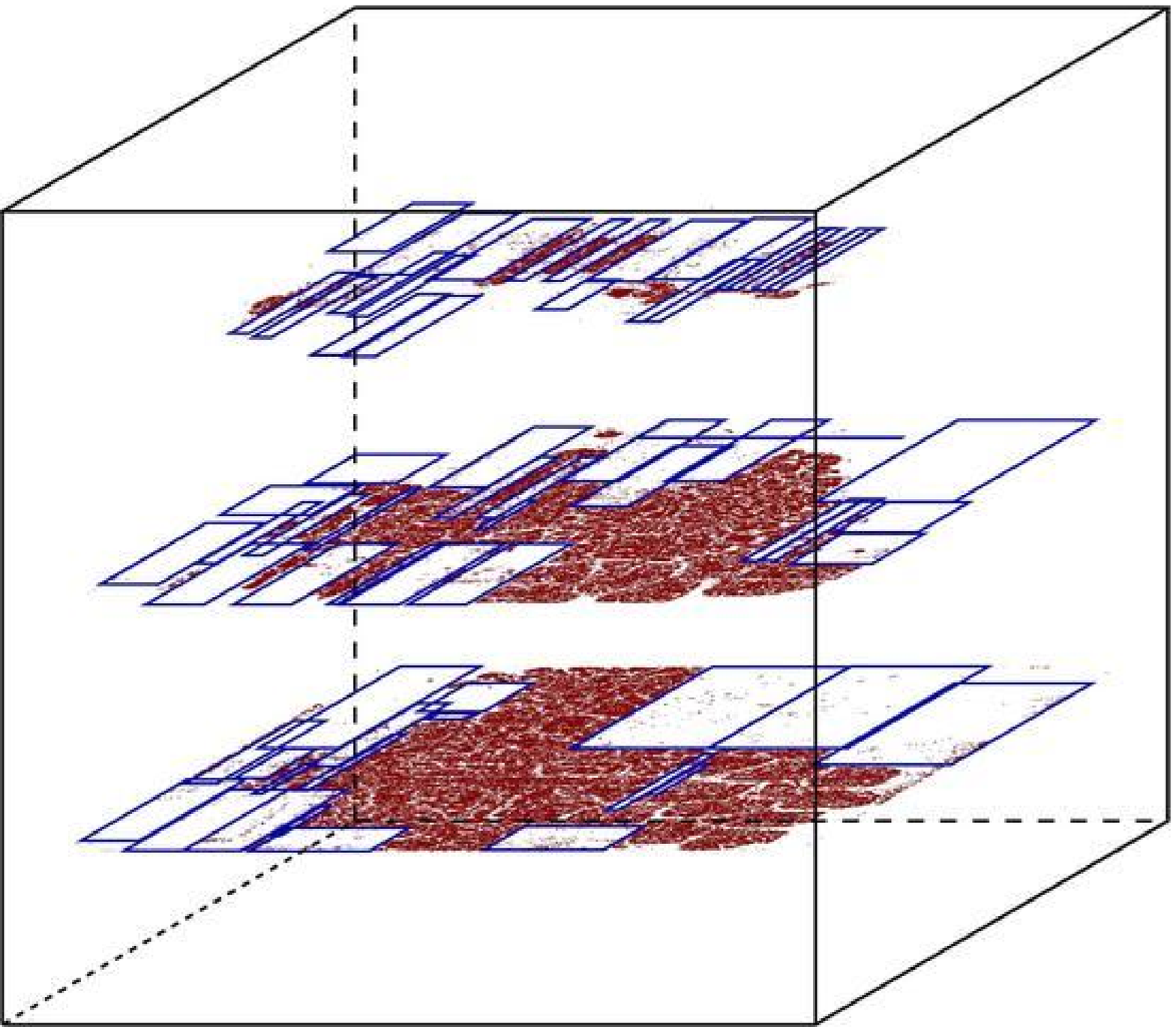}
    \caption{3D macro LG: iter=0, WL=4.99e7, $\#$VI=7.96e3, Om=9.05e5.}
    \label{subfig:den}
  \end{subfigure}
  \begin{subfigure}[b]{0.22\textwidth}
    \centering
    \includegraphics[keepaspectratio, width=0.95\textwidth]{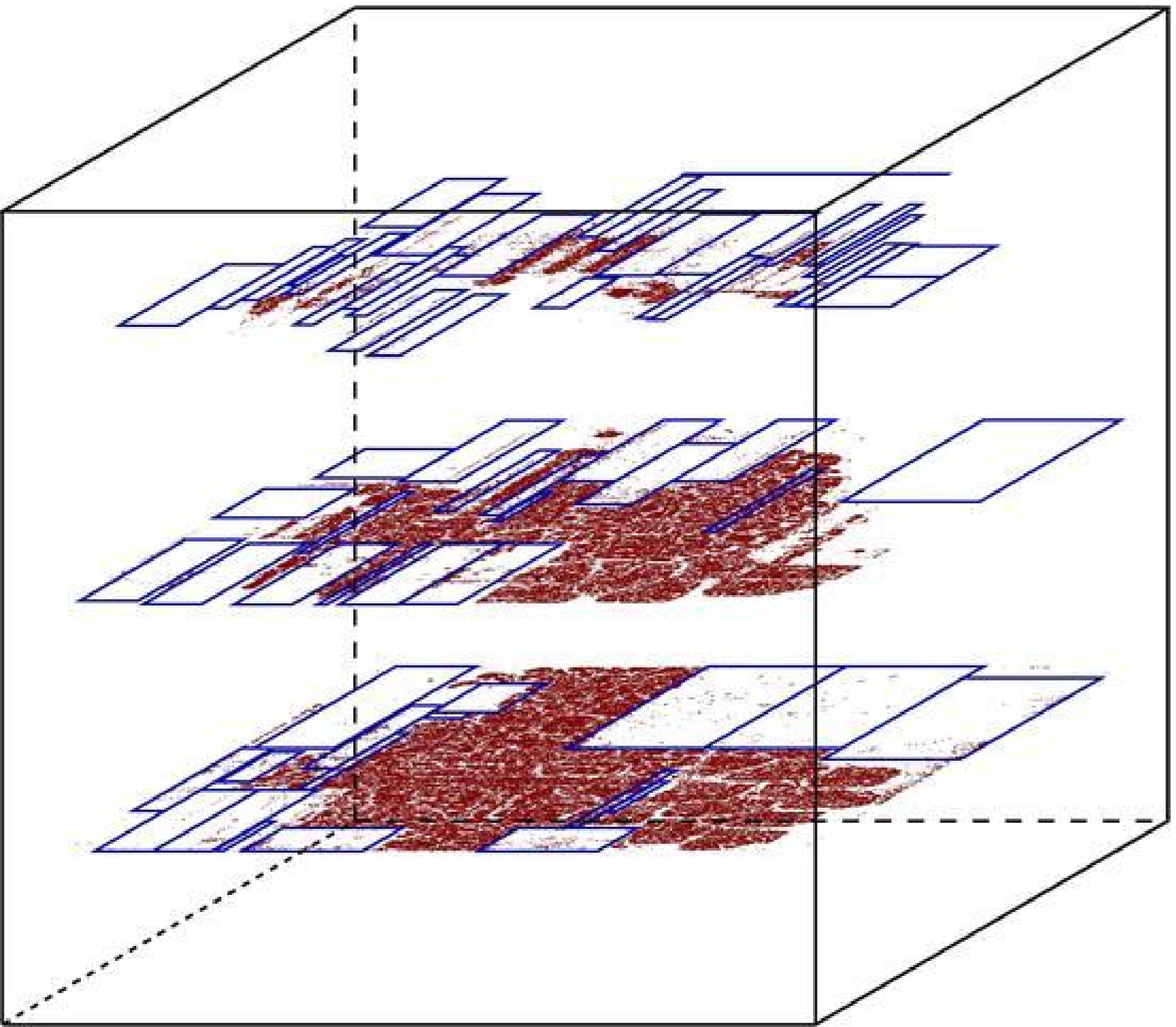}
    \caption{3D macro LG: iter=4, WL=5.10e7, $\#$VI=9.10e3, Om=0.}
    \label{subfig:den}
  \end{subfigure}
  \begin{subfigure}[b]{0.22\textwidth}
    \centering
    \includegraphics[keepaspectratio, width=0.95\textwidth]{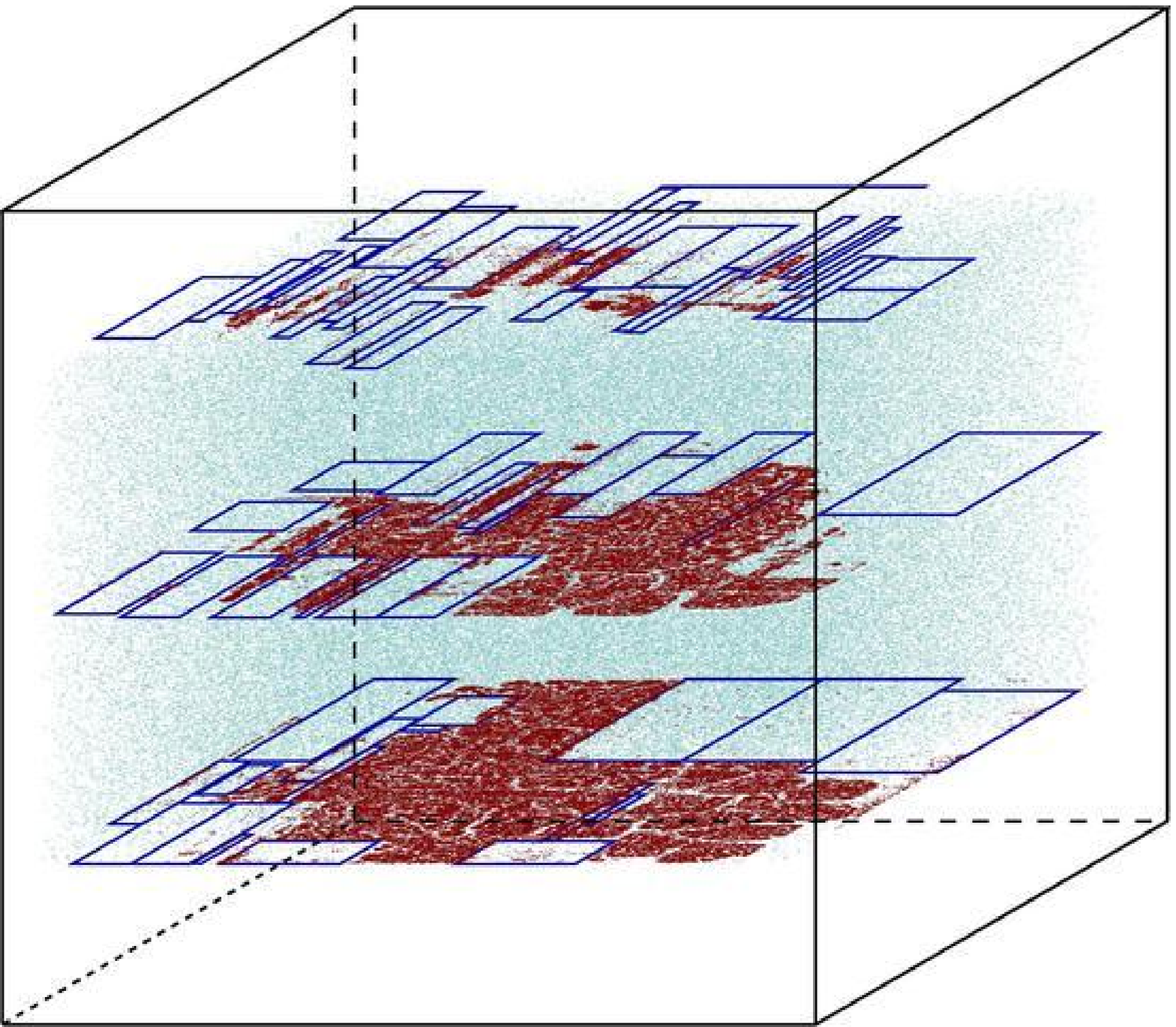}
    \caption{3D standard-cell GP: iter=0, WL=5.10e7, $\#$VI=9.10e3, $\tau=8.1\%$.}
    \label{subfig:den}
  \end{subfigure}
  \begin{subfigure}[b]{0.22\textwidth}
    \centering
    \includegraphics[keepaspectratio, width=0.95\textwidth]{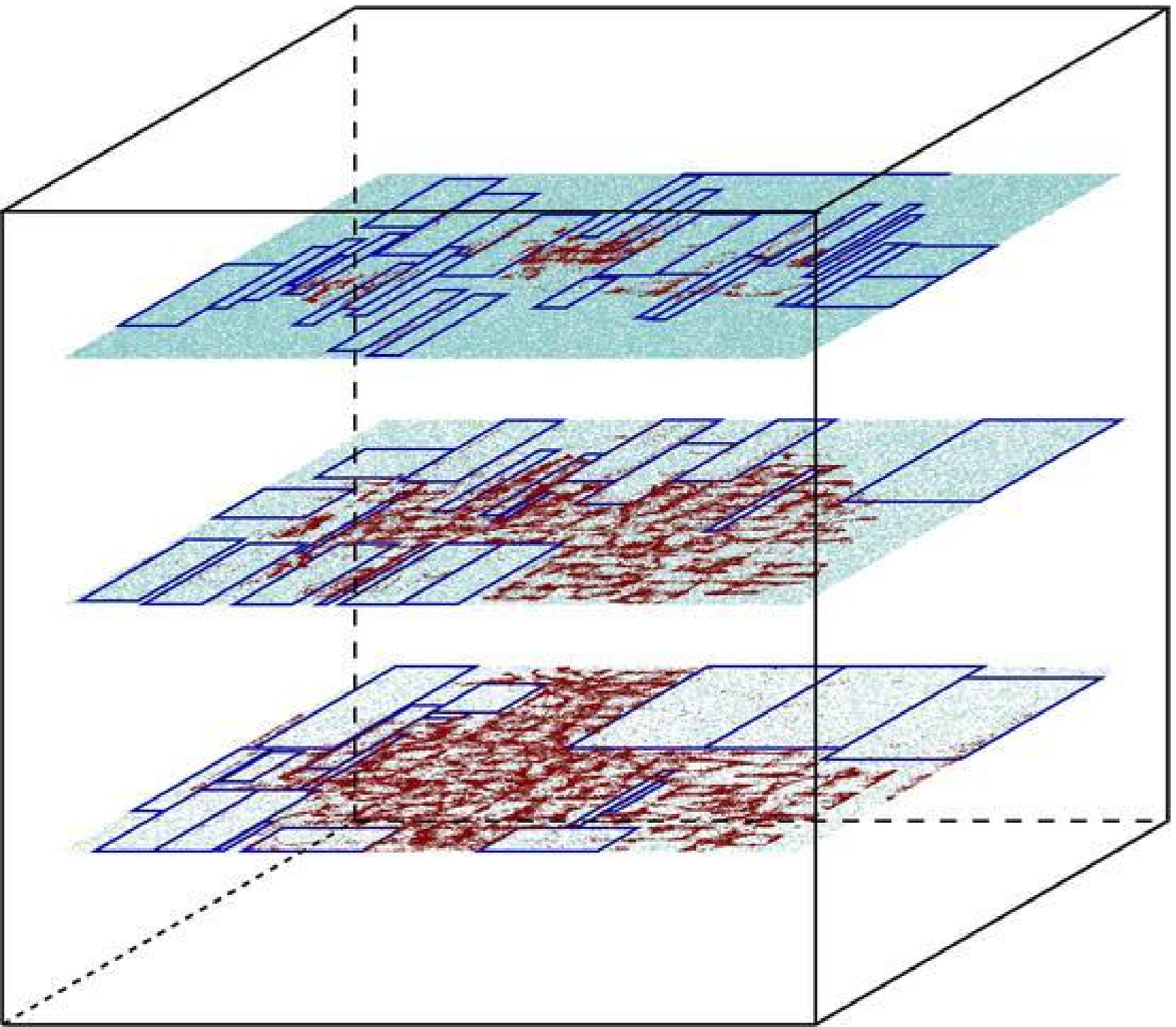}
    \caption{2D standard cell GP: iter=0, WL=4.92e7, $\#$VI=9.10e3, $\tau=66.7\%$.}
    \label{subfig:den}
  \end{subfigure}
  \begin{subfigure}[b]{0.22\textwidth}
    \centering
    \includegraphics[keepaspectratio, width=0.95\textwidth]{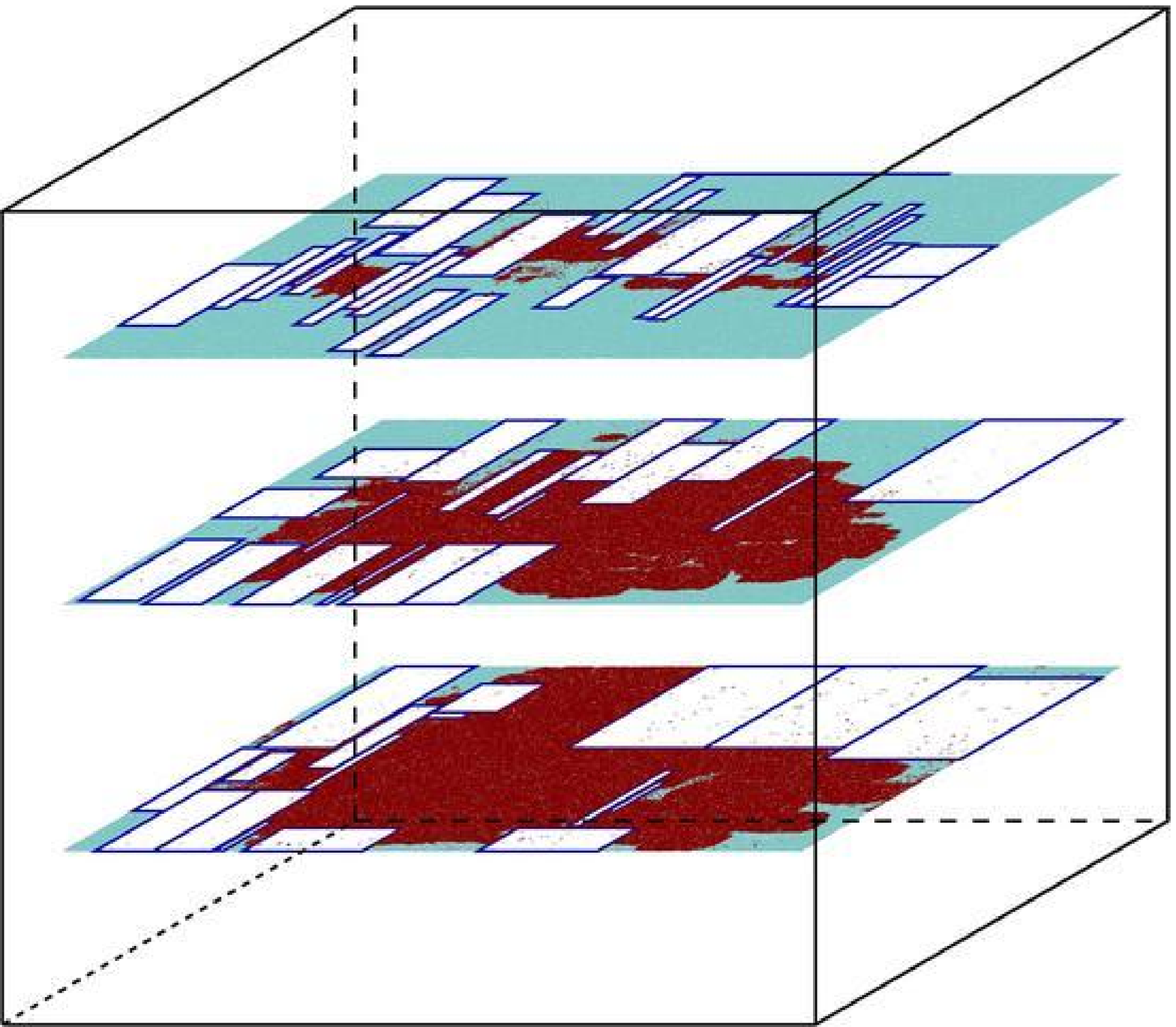}
    \caption{2D standard cell GP: iter=394, WL=5.08e7, $\#$VI=9.11e3, $\tau=14.8\%$.}
    \label{subfig:den}
  \end{subfigure}
  \begin{subfigure}[b]{0.222\textwidth}
    \centering
    \includegraphics[keepaspectratio, width=0.95\textwidth]{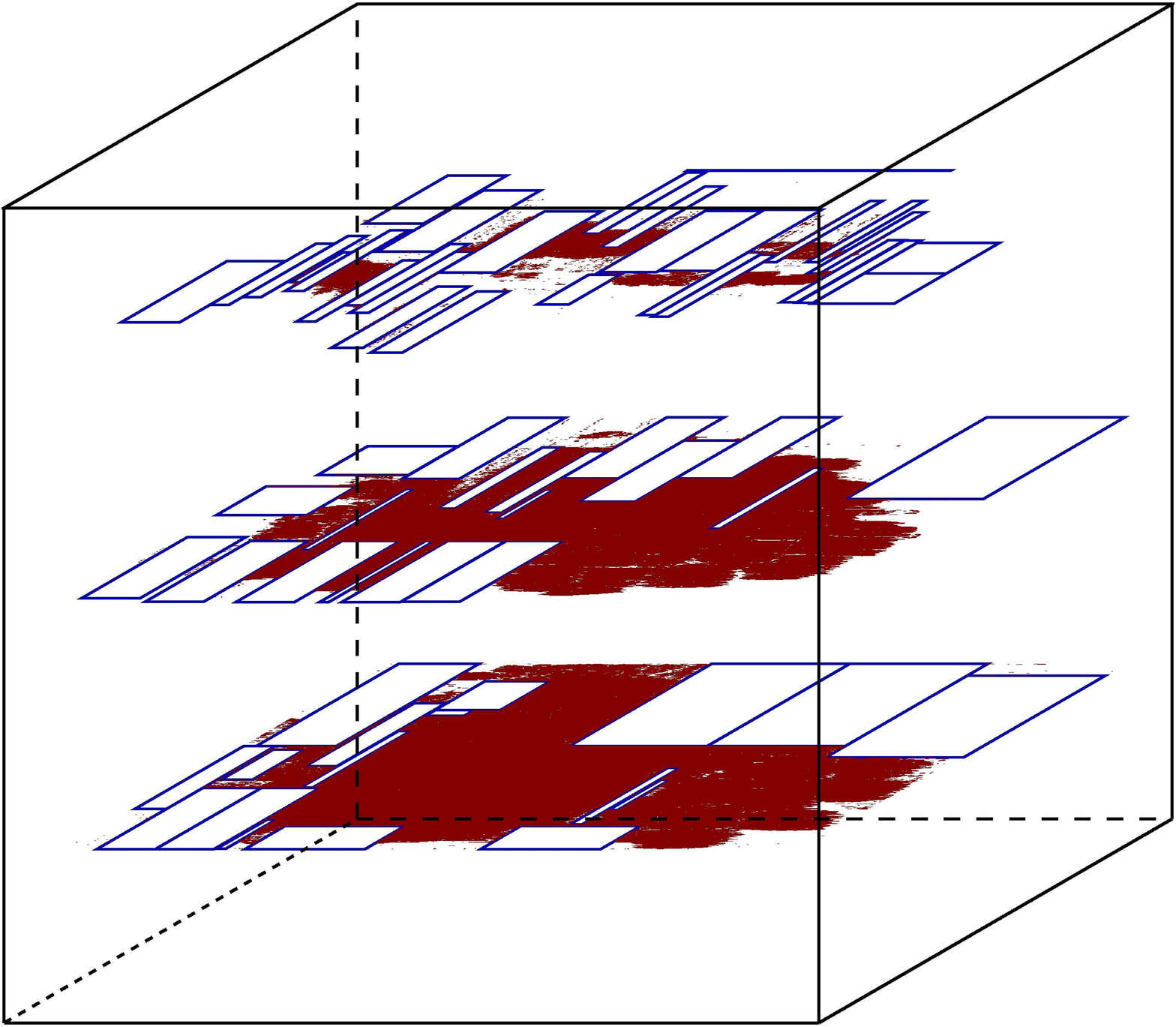}
    \caption{2D standard cell DP: WL=5.42e7, $\#$VI=9.10e3, $\tau=0\%$.}
    \label{subfig:den}
  \end{subfigure}
  \caption{Post-placement on MMS 
  ADAPTEC1 with three tiers. Standard cells, macros and fillers are denoted by red dots, 
  blue rectangles and cyan dots. Om denotes the total macro overlap.}
  \label{fig:mlg3d}
\end{figure}

Our 3D-IC macro legalizer generates a legal 
macro layout with zero macro overlap and small wirelength overhead. 
The algorithm is stochastic based on simulated annealing~\cite{sa83}.
A 3D-IC standard-cell global placement follows to 
mitigate the quality loss due to sub-optimal macro legalization.
We assign standard cells to their closest tiers and conduct 
simultaneous 2D-IC standard-cell placement on all the tiers. 
The standard-cell layouts of all the tiers are locally refined. 
Figure~\ref{fig:mlg3d} shows the respective placement progressions, 
more details can be found in~\cite{eplace-ms}.
The detailed placer from~\cite{fastdp} is then invoked 
for a tier-by-tier standard-cell legalization and detailed placement
from the bottom to the top tier.

\begin{table*}
\centering
\caption{HPWL (e7), $\#$VI (vertical interconnect) (e3) and runtime (minutes) on the IBM-PLACE benchmark suite~\cite{ibm_place}. 
Cited results are marked with $^*$. All the experiments are conducted under single-thread mode. 
The results are evaluated by the same scripts and normalized to ePlace-3D. The best result for each case is in bold-face.}
\begin{tabular}{|c|r|r|r|r|r|r|r|r|r|r|r|}
\hline
\multicolumn{3}{|c|}{Categories}   &
\multicolumn{3}{|c|}{NTUplace3-3D~\cite{wa_tcad}}    & 
\multicolumn{3}{|c|}{mPL6-3D$^*$~\cite{mpl3d_tcad13}} &
\multicolumn{3}{|c|}{ePlace-3D}    \\ \cline{1-12}
\multicolumn{1}{|c|}{Benchmarks}   &
\multicolumn{1}{|c|}{$\#$ Cells}   &
\multicolumn{1}{|c|}{$\#$ Nets}    &
\multicolumn{1}{|c|}{HPWL}         &
\multicolumn{1}{|c|}{$\#$VI}       & 
\multicolumn{1}{|c|}{CPU}          & 
\multicolumn{1}{|c|}{HPWL}         &
\multicolumn{1}{|c|}{$\#$VI}       & 
\multicolumn{1}{|c|}{CPU}          & 
\multicolumn{1}{|c|}{HPWL}         &
\multicolumn{1}{|c|}{$\#$VI}       & 
\multicolumn{1}{|c|}{CPU}          \\ \hline 
IBM01	& 12K	& 12K	& 0.34	& {\bf 0.69}	& 0.20	& 0.26	& 1.04	& 2.95	& {\bf 0.25}	& 1.31	& 0.58  \\ \hline  
IBM03	& 22K	& 22K	& 0.76	& 3.32	& 0.50	& 0.59	& {\bf 3.11}	& 4.72	& {\bf 0.56}	& 3.27	& 1.33  \\ \hline  
IBM04	& 27K	& 26K	& 1.00	& {\bf 2.60}	& 0.60	& 0.81	& 2.95	& 6.41	& {\bf 0.74}	& 3.53	& 1.88  \\ \hline  
IBM06	& 32K	& 33K	& 1.30	& 3.99	& 0.80	& 1.05	& {\bf 3.97}	& 6.20	& {\bf 0.92}	& 4.50	& 2.98  \\ \hline  
IBM07	& 45K	& 44K	& 1.92	& 5.73	& 1.30	& 1.59	& 4.68	& 8.64	& {\bf 1.50}	& {\bf 4.39}	& 3.87  \\ \hline  
IBM08	& 51K	& 48K	& 2.08	& 4.90	& 1.70	& 1.71	& {\bf 3.94}	& 11.23	& {\bf 1.54}	& 4.90	& 4.75  \\ \hline  
IBM09	& 52K	& 50K	& 1.92	& 3.88	& 1.50	& 1.45	& 3.24	& 14.61	& {\bf 1.40}	& {\bf 3.18}	& 5.63  \\ \hline  
IBM13	& 82K	& 84K	& 3.69	& {\bf 3.98}	& 2.60	& 2.88	& 5.59	& 19.62	& {\bf 2.67}	& 4.73	& 8.65  \\ \hline  
IBM15	& 158K	& 161K	& 9.16	& 15.67	& 7.20	& 6.79	& 10.52	& 46.82	& {\bf 6.39}	& {\bf 9.16}	& 40.25 \\ \hline   
IBM18	& 210K	& 201K	& 13.41	& 12.19	& 13.60	& {\bf 9.16}	& 15.22	& 52.09	& 9.47	& {\bf 6.83}	& 63.07 \\ \hline   
\multicolumn{1}{|c|}{Avg.} & 69K & 68K & 
$37.15\%$ & $10.27\%$ & $0.30\times$ & 
$ 6.44\%$ & $ 9.11\%$ & $2.55\times$ & 
$ 0.00\%$ & $ 0.00\%$ & $1.00\times$ \\ \hline
\end{tabular}
\label{tab:res_ibm}
\end{table*}

In general, we have fine-grained 2D 
placement interleaved with coarse-grained 3D placement, 
which achieves a good trade-off between quality and efficiency.
On average of all the ten IBM-PLACE circuits, 
the application of 2D refinement reduces the wirelength by 
more than $4\%$.


\section{Experiments and Results}
\label{sec:exp}

We implement ePlace-3D using C programming 
language in the single-thread mode and execute the program in a 
Linux machine with Intel i7 920 2.67GHz CPU and 12GB memory. 
There is no benchmark specific parameter tuning in our work. 
$\#$VI are controlled by the weighting factor $\beta_z$ based on 
capacitance ratio.
By~\cite{cicc13}, one TSV (VI) has the capacitance of $C_{VI}=30fF$ 
at 45nm tech-node.
ITRS annual reports~\cite{itrs} show that  
unit capacitance of interconnects at intermediate routing layers 
is constantly $2pF/cm$ across various tech-nodes. 
Placement row height is $1.4um$ at 45nm tech-node
($70nm$ M1 half-pitch, ten M1 tracks per row), 
capacitance becomes $C_{ROW}=0.3fF$ for 2D interconnect spanning 
one-row height. 
Based on the length units for each benchmark, as well as our 
geometric transformation of the placement core region to be 
$[0,1]\times[0,1]\times[0,1]$ as discussed in Section~\ref{subsec:bal},
we compute the respective capacitance ratio of one VI versus 
one unit wirelength and use it as the VI weight.
Specifically, we have 
\begin{equation}
\beta_z=\frac{\#tiers\times C_{VI}}{\#rows\times C_{ROW}}\\
\end{equation}

Notice that the focus of this work is the algorithm framework 
of 3D placement, not the accurate weight modeling of vertical connects.
The weighting factor can be adjusted by VLSI 
designers for their particular needs, e.g., 
vertical connects of different electric and physical 
attributes (TSVs, MIVs, super contacts, etc.).

We conduct experiments on 
IBM-PLACE~\cite{ibm_place} standard-cell benchmarks without 
macros or blockages, 
all of which are derived from real IC design.
We include two state-of-the-art 3D-IC placers, 
mPL6-3D~\cite{mpl3d_tcad13}\footnote{Although mPL6-3D has 
extension to thermal-aware placement, its experiments on 
the IBM-PLACE cases are based on their original 
prototype driven by only wirelength and density but not thermal.}, 
and NTUplace3-3D~\cite{wa_tcad}, in our experiments on IBM-PLACE.
As other categories of algorithms (e.g., folding and partition based 
approaches) have been outperformed by analytic placement in literature, 
we do not include them in our experiments.
We have obtained the binary of NTUplace3-3D from the original authors 
and executed it on our machine for experiments\footnote{There 
is a small quality gap on NTUplace3-3D between our local experiment 
results and that published in~\cite{wa_tcad}, which may be due to the 
differences in computing platforms.}.
mPL6-3D is not available (as notified by the author), 
so we cite the performance from their latest publication~\cite{mpl3d_tcad13}.
We use exactly the same benchmark transformation as that by mPL6-3D and NTUplace3-3D. 
I.e., we insert four silicon tiers into each benchmark, 
scale down each tier to $\frac{1}{4}$ of the original 2D placement area, 
add $10\%$ whitespace to each tier, and keep the aspect ratio of 
each tier to be the same as the original 2D design.
As a result, 
all the experiments on the three placers, 
including those from~\cite{mpl3d_tcad13}, 
are conducted on exactly the same {\bf IBM-PLACE-3D} benchmarks. 
As HPWL and $\#$VI are being computed in exactly the same way, 
the performance comparison among the three placers are fair. 
The results on IBM-PLACE cases are shown in Table~\ref{tab:res_ibm}.
On average of all the ten circuits, ePlace-3D 
outperforms mPL6-3D and NTUplace3-3D with 
$6.44\%$ and $37.15\%$ shorter wirelength 
together with 
$9.11\%$ and $10.27\%$ fewer VIs. 
ePlace-3D runs $2.55\times$ faster than mPL6-3D but 
$0.30\times$ slower than NTUplace3-3D, nevertheless, 
the improvement on wirelength ($37.15\%$) and 
VI ($10.27\%$) is significant.

\begin{table*}
\begin{small}
\centering
\caption{HPWL (e6), $\#$VI (vertical interconnect) and runtime (mins) on MMS circuits.
Cited results are marked with $^*$. All the experiments are in single-thread mode.
The HPWL and CPU results are normalized to the best published 2D placement results~\cite{eplace-ms}, 
$\#$VI are normalized to $\#$ objects.}
\begin{tabular}{|c|r|r|r|r|r|r|r|r|r|r|r|r|r|}
\hline
\multicolumn{3}{|c|}{$\#$ tiers}   &
\multicolumn{2}{|c|}{ePlace-MS$^*$~\cite{eplace-ms}} &
\multicolumn{3}{|c|}{ePlace-3D w/ 2 tiers} & 
\multicolumn{3}{|c|}{ePlace-3D w/ 3 tiers} &
\multicolumn{3}{|c|}{ePlace-3D w/ 4 tiers} \\ \cline{1-14}
\multicolumn{1}{|c|}{Benchmarks}   &
\multicolumn{1}{|c|}{$\#$ Objs}    &
\multicolumn{1}{|c|}{$\#$ Nets}    & 
\multicolumn{1}{|c|}{HPWL}         & 
\multicolumn{1}{|c|}{CPU}          &
\multicolumn{1}{|c|}{HPWL}         & 
\multicolumn{1}{|c|}{$\#$VI}       &
\multicolumn{1}{|c|}{CPU}          &
\multicolumn{1}{|c|}{HPWL}         & 
\multicolumn{1}{|c|}{$\#$VI}       &
\multicolumn{1}{|c|}{CPU}          &
\multicolumn{1}{|c|}{HPWL}         & 
\multicolumn{1}{|c|}{$\#$VI}       &
\multicolumn{1}{|c|}{CPU}          \\ \hline
ADAPTEC1& 	211K	& 221K	& 67.15		& 5.47		& 59.51		& 5733	& 24.63	&	54.19	&9104	&14.65	&51.3	&13568	&16.03  \\ \hline  
ADAPTEC2& 	255K	& 266K	& 77.37		& 7.43		& 73.97		& 9269	& 39.67	&	75.38	&9929	&25.18	&59.97	&18085	&24.57  \\ \hline  
ADAPTEC3& 	451K	& 466K	& 164.50	& 27.23		& 141.97	& 5557	& 95.48	&	136.85	&18203	&88.55	&120.29	&28694	&94.42  \\ \hline  
ADAPTEC4& 	496K	& 515K	& 148.38	& 29.35		& 126.94	& 8149	& 107.15&	113.22	&13811	&121.40	&106.34	&14527	&118.13 \\ \hline   
BIGBLUE1& 	278K	& 284K	& 86.82		& 7.82		& 76.06		& 8272	& 40.63	&	71.34	&10508	&36.17	&63.64	&19403	&38.05  \\ \hline  
BIGBLUE2& 	557K	& 577K	& 130.18	& 13.70		& 109.27	& 2565	& 70.25	&	97.1	&5347	&63.58	&90.14	&9241	&64.95  \\ \hline  
BIGBLUE3& 	1096K	& 1123K	& 302.29	& 72.98		& 251.77	& 24466	& 268.47&	271.27	&42053	&291.38	&295.38	&62669	&388.08 \\ \hline   
BIGBLUE4& 	2177K	& 2229K	& 657.92	& 204.15	& 577.98	& 21263	& 491.97&	537.2	&50552	&563.98	&500.25	&113590	&420.17 \\ \hline   
ADAPTEC5& 	843K	& 867K	& 310.54	& 48.35		& 258.18	& 22705	& 170.90&	244.57	&27764	&146.22	&223.44	&50732	&149.22 \\ \hline   
NEWBLUE1& 	330K	& 338K	& 61.85		& 10.87		& 56.36		& 5901	& 28.15	&	53.05	&7295	&24.08	&48.85	&12346	&25.07  \\ \hline  
NEWBLUE2& 	441K	& 465K	& 162.93	& 62.40		& 179.82	& 25571	& 67.27	&	143.5	&43642	&77.20	&169.78	&53487	&72.98  \\ \hline  
NEWBLUE3& 	494K	& 552K	& 304.15	& 17.53		& 240.47	& 7686 	&308.62	&	365.10	&48979	&410.73	&397.46	&51597	&265.67	\\ \hline    
NEWBLUE4& 	646K	& 637K	& 228.54	& 29.73		& 197.21	& 11372	& 110.02&	177.82	&29767	&112.80	&171.21	&35067	&101.78 \\ \hline   
NEWBLUE5& 	1233K	& 1284K	& 392.27	& 63.40		& 344.95	& 45995	& 202.12&	303.05	&64336	&195.52	&280.42	&95768	&216.22 \\ \hline   
NEWBLUE6& 	1255K	& 1288K	& 408.36	& 69.65		& 379.59	& 10901	& 222.72&	325.35	&50487	&194.57	&298.82	&66983	&180.88 \\ \hline   
NEWBLUE7& 	2507K	& 2636K	& 894.31	& 191.47	& 814.79	& 18615	& 363.30&	696.27	&92943	&375.65	&670.51	&111562	&353.92 \\ \hline   
\multicolumn{1}{|c|}{ Avg.} & 829K & 859K &
$ 0.00\%$ & $1.00\times$ & 
$13.67\%$ & $2.17\%$  & $3.13\times$ & 
$20.50\%$ & $4.30\%$  & $3.03\times$ & 
$27.54\%$ & $6.10\%$  & $2.94\times$ \\ \hline
\end{tabular}
\end{small}
\label{tab:res_mms}
\end{table*}

To validate the scalability of ePlace-3D, 
we also conduct experiments on the large-scale 
{\bf modern mixed-size (MMS)} benchmarks~\cite{flop}
with on average 829K and up to 2.5M netlist objects. 
MMS benchmarks was first published in DAC 2009. 
The circuits inherit the same netlists and density 
constraints $\rho_t$ from ISPD 2005~\cite{ispd05} and 
ISPD 2006~\cite{ispd06} benchmarks but have 
all the macros freed to place. The original 
planar placement domain is geometrically transformed to be 
of $2$, $3$ and $4$ silicon tiers, each tier is equally downsized 
to keep both the aspect ratio and total silicon area unchanged. 
All the standard cells and macros keep their original 
dimensions and span only one tier.
MMS circuits have all their fixed objects with zero area (volume) 
and outside the placement boundaries, 
and we geometrically transform them to the boundary of the 
bottom (first) tier.
Also, as macros are all free to move, 
we skip the geometrical transformation of the fixed macro layout 
from 2D to 3D, which is sub-optimal and usually causes quality loss.
Similar to mPL6-3D~\cite{mpl3d_tcad13} and NTUplace3-3D~\cite{wa_tcad}, 
we add $10\%$ extra whitespace to each tier, in order to 
relieve the placement dilemma due to the increased 
area ratio between large macros and silicon tiers\footnote{
BIGBLUE3, NEWBLUE2 and NEWBLUE3 have very large macros. 
For the tier insertion of two, three and four, 
we add $20\%$, $30\%$ and $40\%$ whitespace to each tier to 
make sure that the largest macro can be accommodated.}.
There are benchmark-dependent target density $\rho_t$ 
for eight out of the sixteen MMS circuits. 
Detailed circuit statistics can be 
found in Table 1 of~\cite{flop}.
We create evaluation scripts to compute the total 
wirelength, number of vertical interconnects, 
and legality 
of the produced 3D-IC placement solution. 
The results on the MMS benchmarks are shown in Table 2. 
Notice that here HPWL is the original half-perimeter wirelength.
It is not penalized by the amount of density overflow, 
since the density overflow in 3D domain is of one more 
dimension thus hard to compare with that of 2D domain.
The binary of NTUplace3-3D does not work with these benchmarks, 
while the binary of mPL6-3D is not available for use.
As a result, we compare the 3D MMS placement solutions with 
the best published (golden) 2D results in literature~\cite{eplace-ms}.
By using two, three and four tiers, 
ePlace-3D outperforms the golden 2D placement  
with on average $13.67\%$, $20.50\%$ and $27.54\%$ shorter wirelength.
On the other side, the average ratio between 
the number of vertical interconnect units versus
the number of placement objects (standard cells and macros)
are only $2.17\%$, $4.30\%$ and $6.10\%$, respectively. 
These vertical connect ratios are much smaller than the average VI ratio on IBM-PLACE, 
which are more than $9\%$ for all the three placers in Table~\ref{tab:res_ibm}.
Due to the introduction of the third dimension, the search space of placement optimization 
is substantially enlarged. However, the runtime increase is just $3\times$, which indicates 
high efficiency of ePlace-3D.

\begin{figure}[http]
\centering
\includegraphics[width=1.0\columnwidth, angle=0]{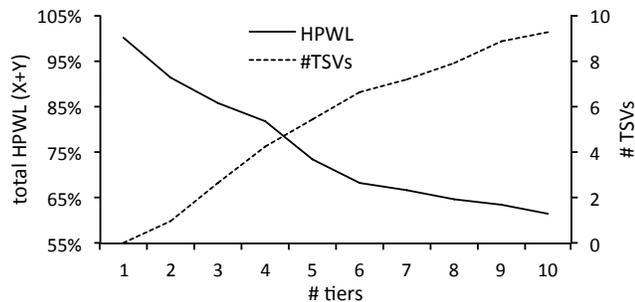}
\caption{Avg. HPWL and $\#$VI of eight selected MMS cases w.r.t. number of silicon tiers.}
\label{fig:multi_tiers}
\end{figure}

\begin{figure}[http]
\centering
\includegraphics[width=1.0\columnwidth, angle=0]{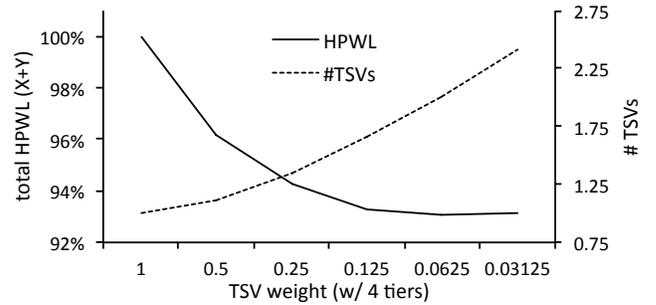}
\caption{Avg. HPWL and $\#$VI of all the sixteen MMS cases w.r.t. VI weights.}
\label{fig:multi_weight}
\end{figure}

We also study the trends of HPWL and $\#$VI by 
linearly sweeping the number of tiers and exponentially 
sweeping the VI weight. We select eight out of the sixteen MMS 
benchmarks (ADAPTEC1, ADAPTEC4, BIGBLUE1, BIGBLUE2, BIGBLUE3, 
BIGBLUE4, NEWBLUE6, NEWBLUE7), all of which could accommodate the 
maximum macro block after inserting ten tiers. Keeping the same 
aspect ratio, the area of each tier is scaled down by ten times 
with the insertion of $10\%$ extra whitespace.
Figure~\ref{fig:multi_tiers} shows that ePlace-3D is able to 
reduce the total 2D wirelength by up to $40\%$ (with the insertion 
of up to ten tiers), while $\#$VI is roughly scaled up by the number of tiers.
VI weight sweeping is conducted on all the sixteen MMS 
benchmarks. 
Figure~\ref{fig:multi_weight} shows the trends of average 
HPWL and $\#$VI by dividing the normal VI weight by up to 
32 times (i.e. $\times$ 0.03125).
The total 2D HPWL saturates at the reduction of $7\%$, while 
$\#$VI is scaled up by roughtly $2.5\times$.

Our 3D-IC placement algorithm shows significant quality 
improvement while limited runtime overhead. 
BIGBLUE4 and NEWBLUE7 are the largest circuits with 2.2M and 2.5M cells, 
and they consume the longest runtime on the 3D-IC placement.
However, compared to the respective golden 2D placement 
solutions, the runtime ratio is upper-bounded by $2.5\times$, 
which is still less than the average runtime ratios of 
$3.13\times$ for two tiers, 
$3.03\times$ for three tiers and $2.94\times$ for four tiers, 
respectively.
To this end, 
ePlace-3D shows good scalability and 
acceptable efficiency on the large cases.

In this work, we do not test ePlace-3D on circuits with 
fixed macros, 
as geometrically transforming the 2D floorplan into 3D is 
difficult and usually error-prone. 
However, ePlace-3D shows high performance and scalability 
on MMS benchmarks with lots of movable large macros, which is 
more difficult to place than fixed-macro layouts.
As a result, we are confident on the performance of ePlace-3D on 
any circuits with fixed macros.
The advantage of 3D tier insertion vanishes if there are 
large macros to accommodate (BIGBLUE3, NEWBLUE3, etc.).
Transformation of 2D planar macros into 3D cuboid macros would resolve 
this issue and ensure the consistent benefits by inserting more tiers.
However, it is beyond this work and will be covered in future.

\section{Conclusion}
\label{sec:conc}

We propose the first electrostatics based placement algorithm {\bf ePlace-3D}, 
which is effective and efficient for 3D-ICs with
uniform exploration over the entire 3D space.
Our 3D-IC density function leverages the analogy between 
placement spreading and electrostatic equilibrium, 
while global and uniform smoothness is realized 
at all the three dimensions.
Our balancing and preconditioning techniques 
prevent solution oscillation or divergence. 
The interleaved 3D coarse-grained optimization followed by 
2D fine-grained post processing obtains a 
good trade-off between quality and efficiency.
The experimental results validate the high performance and 
scalability of our approach, 
indicating the benefits of placement smoothness. 
In future, we will develop 3D density function to 
address the volume of vertical interconnects (VI).
We would also like to explore advanced technology for 3D-IC 
placement/routing with patterning and 
graph coloring technology~\cite{lin13}.

\section{Acknowledgment}
\label{sec:ack}
The authors acknowledge 
(1) Prof. Dae Hyun Kim and Prof. Sung Kyu Lim for 
providing the 3D-IC flow scripts and IWLS testcases
(2) Dr. Meng-Kai Hsu and Prof. Yao-Wen Chang for 
providing the binary of NTUplace3-3D 
(3) Dr. Guojie Luo and Prof. Jason Cong for providing 
the binary of mPL6-3D
(4) the support of NSF CCF-1017864.

\scriptsize

\bibliographystyle{abbrv}
\bibliography{pl}

\end{document}